\documentclass[journal]{IEEEtran}
\RequirePackage{graphicx}
\ifCLASSINFOpdf
  \usepackage{graphicx}
  \usepackage{subcaption}
  \usepackage{tikz}

\else

\fi

\usepackage{amsmath}
\usepackage{amssymb}
\usepackage{hhline}
\usepackage{amsthm}      

\usepackage{pifont}

\usepackage{algorithm}
\usepackage{algpseudocode}
\usepackage{multirow}

\usepackage[compact]{titlesec}
\titlespacing{\section}{0pt}{*0.7}{*0.3}
\titlespacing{\subsection}{0pt}{*0.5}{*0.3}
\titlespacing{\subsubsection}{0pt}{*0.3}{*0.2}

\usepackage{enumitem}
\usepackage{caption}

\usepackage{makecell}
\usepackage{array}

\usepackage{subcaption}
\IEEEoverridecommandlockouts
\usepackage{cite}
\usepackage{amsfonts}
\usepackage{textcomp}
\usepackage{xcolor}
\def\BibTeX{{\rm B\kern-.05em{\sc i\kern-.025em b}\kern-.08em
    T\kern-.1667em\lower.7ex\hbox{E}\kern-.125emX}}

\begin{document}
%
\title{\Huge Efficient Graph Neural Networks for Multicarrier Wideband Hybrid Beamforming Optimization
\thanks{This work was supported in part by the National Science Foundation under Grant 2234122 and Grant 2403285. A preliminary result of the work in this paper was presented at the 2024 IEEE International Symposium on Phased Array Systems and Technology, October 15–18, 2024 [DOI: 10.1109/ARRAY58370.2024.10880428].

The authors are with the Department of Electrical and Computer Engineering,
Tufts University, Medford, MA 02155 USA (e-mail: Beier.Li@tufts.edu;
Mai.Vu@tufts.edu).

This article has been published in \textit{IEEE Transactions on Wireless Communications}. DOI: 10.1109/TWC.2026.3732590.

\copyright~2026 IEEE. Personal use of this material is permitted.  Permission from IEEE must be obtained for all other uses, in any current or future media, including reprinting/republishing this material for advertising or promotional purposes, creating new collective works, for resale or redistribution to servers or lists, or reuse of any copyrighted component of this work in other works.}
}

\author{Beier~Li,~\IEEEmembership{Student~Member,~IEEE,}        Mai~Vu,~\IEEEmembership{Senior Member,~IEEE}}

\maketitle

\begin{abstract}
6G wireless technology is poised to adopt higher and wider frequency bands, leveraging highly directional beamforming. However, the vast bandwidths amplify the impact of beam squinting. Traditional solutions, such as adding a true-time-delay filter to each antenna, are cost-prohibitive due to the required hardware scale. This paper proposes a signal processing alternative using Graph Neural Networks (GNNs) to optimize hybrid beamforming in multicarrier wideband systems. Using a bipartite graph to represent a shared analog beamformer among multiple subcarriers, we develop three GNN structures with distinct digital beamformer representations (i) at the subcarrier nodes, (ii) at the edges, or (iii) integrating traditional singular-value decomposition solutions. By designing an efficient message-passing mechanism, these structures offer insights into the impact of different GNN designs on communication system performance and computational complexity. Extensive analysis and ablation studies show that our proposed GNN structures outperform traditional optimization methods and existing ML-based solutions. Furthermore, the proposed GNNs exhibit strong resiliency to beam squinting and better robustness against imperfect CSI than even fully digital beamforming and all existing hybrid designs. These GNNs can also be extended to multi-user scenarios and demonstrate excellent generalization capabilities, allowing trained models to adapt to diverse multicarrier and multi-user settings without retraining.
\end{abstract}


%
\IEEEpeerreviewmaketitle

\section{Introduction}
\IEEEPARstart{H}{ybrid} beamforming, which combines analog and digital techniques, offers a cost-effective solution and robust performance for employing massive antenna arrays in modern communication systems. In wideband systems utilizing OFDM to enhance data rates and resistance to multipaths, the beam pattern increasingly varies with frequency across subcarriers \cite{beamsquint}. This phenomenon, known as beam squinting, becomes particularly significant in 6G wireless networks operating in the sub-Terahertz (THz) spectrum from 100 GHz to 1 THz, where the bandwidth can be as wide as 18 GHz \cite{THz}.

To manage beam squinting in hybrid beamforming systems, recent research efforts have focused on two main approaches: true-time-delay lines (TTD) and signal processing methods. TTD is a time-delay filter integrated into each antenna to provide precise control over signal timing and effectively eliminates beam squinting \cite{TTD_subarray}. However, in the sub-THz frequency range, the ability to pack more antennas into the same device size not only enhances performance but also significantly increases the number of TTDs required, leading to substantially higher costs. Given these economic considerations, signal processing methods have become an attractive alternative for managing beam squinting in high-frequency domains because of their cost-efficiency.

Since our focus in this paper is on managing the beam squinting in MIMO-OFDM hybrid beamforming systems, we primarily consider a single-user scenario. This allows us to isolate beam squinting from inter-user interference, which enables the design and evaluation of more effective hybrid beamforming strategies. As a natural extension, we further apply the proposed methods to multi-user scenarios to demonstrate their applicability under more general system settings.

\subsection{Related Work}
Despite recent progress, current literature indicates that beamforming designs for wideband channels lack efficient solutions to effectively address beam squinting \cite{sixmethods}. Optimization-based algorithms such as Alternative Manifold Optimization (AMO) \cite{AMO} and Iterative Coordinate Descent (ICD) \cite{ICD} have shown promise, but suffer from the need for continuous optimization with every channel update and hence require significant computational resources, limiting their practical advantage. In contrast, machine learning (ML)-based approaches \cite{DNN_SU, CNN_beam, LSTMCNN_beam, compare_CNN_GNN, GNN_wireless, Li2025, Wang2024GNN, Yang2024FDD, Shen2022, Huang2024Hetero, Wan2024Scalable} offer a promising alternative that can significantly reduce this burden.


\textit{1) Traditional Optimization-based Methods:} A comparative study has explicitly analyzed beam squint under wideband conditions \cite{sixmethods}, which compares six analog beamformer designs. While computationally efficient, these methods substantially degrade system performance. In the most severe cases, when the number of antennas increases to 160, the performance drops to $70\%$ of its optimal level \cite{sixmethods}.



The benchmark for this comparison was the centralized optimization algorithm Alternative Manifold Optimization (AMO) \cite{AMO}, which uses manifold optimization to calculate the analog beamformer and then applies a pseudo-inverse for digital beamformers. AMO shows strong beam squinting resistance and achieves the best to date spectral efficiency performance among traditional hybrid beamforming methods.

Another optimization method is the Iterative Coordinate Descent (ICD) algorithm \cite{ICD}, which sequentially updates the elements in the analog beamformer one at a time while keeping others fixed. This greedy strategy sacrifices some performance compared to AMO but reduces memory requirements.

{These traditional optimization-based methods have been widely adopted as benchmarks in existing machine learning–based hybrid beamforming studies \cite{DNN_SU, CNN_beam, LSTMCNN_beam, GNN_wireless, Li2025, Wang2024GNN, Yang2024FDD, Shen2022, Huang2024Hetero, Wan2024Scalable}.}

\begin{table*}[t!]
\centering
\footnotesize
\renewcommand{\arraystretch}{1.3}
\caption{Comparison of GNN-based Hybrid Beamforming Methods}
\label{tab:gnn_comparison}
\hspace*{-0.2cm}
{
\begin{tabular}{|
>{\centering\arraybackslash}m{2.1cm}|
>{\centering\arraybackslash}m{5.8cm}|
>{\centering\arraybackslash}m{4.8cm}|
>{\centering\arraybackslash}m{2.0cm}|
>{\centering\arraybackslash}m{1.4cm}|
}
\hline
\textbf{Refs.} & \textbf{Problem Descriptions} & \textbf{Graph Structure} & \textbf{Generalization Ability} & \textbf{Outperform AMO \cite{AMO}?} \\
\hline
\cite{GNN_wireless} & A unified multi-dimensional graph model for beamforming and other wireless tasks. & K-partite graph (nodes can be users, subcarriers, antennas, etc.) & Weak & \ding{55} \\
\hline
\cite{Wang2024GNN} & End-to-end hybrid beamforming & Near fully-connected graph (nodes are base stations, users, and subcarriers) & Moderate & \ding{55} \\
\hline
\cite{Yang2024FDD}  & Joint design of pilot, channel feedback, and hybrid beamforming & Bipartite graph (analog and digital beamformer nodes) & Weak & \ding{55} \\
\hline
\textbf{NU-GNN (Ours)} & \multirow{3}{6.0cm}{\centering Wideband hybrid beamforming, beamsquinting mitigation} & \multirow{3}{4.8cm}{\centering Bipartite graph (analog and subcarrier nodes)} & Strong & \ding{51} \\
\cline{1-1} \cline{4-5}
\textbf{EU-GNN (Ours)} &  &  & Strong  & \ding{51} \\
\cline{1-1} \cline{4-5}
\textbf{AN-GNN (Ours)} &  &  & Strong & \ding{55} \\
\hline
\end{tabular}
}
\vspace{-7pt}
\end{table*}
\textit{2) Machine Learning-based Methods:} Recent ML-based methods can approach the data rate obtained by traditional optimization algorithms while improving computational efficiency. For example, several studies have applied fully-connected neural networks (FNNs) \cite{DNN_SU} and convolutional neural networks (CNNs) \cite{CNN_beam} to hybrid beamforming in single-user MIMO systems, and have also developed several CNN-based methods for multiuser narrowband scenarios \cite{LSTMCNN_beam, compare_CNN_GNN}. However, the method in \cite{LSTMCNN_beam} relies on a pseudo-inverse to form the analog beamformer, making it suboptimal for extension to OFDM systems. A theoretical analysis in \cite{compare_CNN_GNN} shows that beamforming requires modeling global dependencies across the entire channel matrix, which conflicts with the local pattern extraction nature of convolutional kernels. Capturing such global interactions with CNNs requires large kernel sizes and leads to a significant increase in trainable parameters as the system size grows. In contrast, a comparison of GNNs with FNNs and CNNs in wireless communication tasks \cite{GNN_wireless} highlights that GNNs are inherently permutation equivariant and naturally scalable to variable-sized graphs, making them more suitable for systems with dynamic configurations. These properties allow GNNs to outperform other ML architectures while maintaining high inference speed.

{
In the context of hybrid beamforming design, most existing GNN-based methods have shown promising results in narrowband scenarios \cite{Shen2022, Li2025, Huang2024Hetero, Wan2024Scalable}. These methods, however, cannot be directly extended to multicarrier settings such as OFDM systems for various reasons, including the coupling of analog and digital components \cite{Shen2022}, relying on a fully connected graph structure where the number of edges grows quadratically with the number of nodes, leading to high computational complexity \cite{Li2025}, being specifically designed for a system setting of coexisting sub-6GHz and mmWave \cite{Huang2024Hetero}, or being tailored to a partially connected hardware structure \cite{Wan2024Scalable}.
}

For GNN-based approaches in multicarrier settings, existing works typically use a naive message-passing mechanism in which a single MLP generates messages that are linearly combined to update node representations \cite{GNN_wireless, Wang2024GNN, Yang2024FDD}. The hyper-edge-based method in \cite{GNN_wireless} models antennas, users, and subcarriers as distinct node types, replicating information for scalability. However, the lack of problem-specific graph design and naive message-passing strategy limits existing GNNs' generalization ability. Another end-to-end approach \cite{Wang2024GNN} utilizes a near fully-connected graph designed for TDD systems to infer downlink channel state information (CSI) from uplink measurements, but its learning update mechanism limits the overall performance. A related study \cite{Yang2024FDD} uses a GNN to jointly design pilot signals, channel feedback, and hybrid beamforming in an FDD system. When perfect CSI is available, the problem reduces to standard hybrid beamforming, making it a suitable baseline for our work.


\subsection{Contributions}

{In this work, we consider HBF for wideband multicarrier systems with beamsquinting effect, and aim to design GNN structures that not only outperforms traditional optimization methods but also effectively mitigate beam squinting. This is in contrast to existing ML solutions  \cite{Shen2022, Li2025, Huang2024Hetero, Wan2024Scalable, GNN_wireless, Wang2024GNN, Yang2024FDD} which are designed for narrowband systems and typically only approach the spectral efficiency performance of traditional optimization-based methods rather than outperforming them. In addition, beam squinting resiliency was not addressed in prior ML-based designs \cite{GNN_wireless, Wang2024GNN, Yang2024FDD}. To fill this gap, we focus on a wideband, single-user MIMO-OFDM system. See Table~\ref{tab:gnn_comparison} for comparison with existing works.}

Specifically, we propose to solve the highly non-convex hybrid beamforming design problem by constructing three different GNN structures using a bipartite graph. The graph has two types of nodes, which represent the analog beamformer and the subcarriers, respectively. The GNNs employ an efficient yet effective message-passing mechanism to jointly learn the beamformers, with an unsupervised loss function that directly maximizes the system's spectral efficiency. Our major contributions are summarized as follows:
\begin{itemize}[leftmargin=1.3em]
    \item We design the GNNs to directly optimize the phases of the analog beamformer instead of its complex-value entries. This direct phase learning has not been explored in prior work \cite{DNN_SU, CNN_beam, LSTMCNN_beam, compare_CNN_GNN, GNN_wireless, Li2025, Wang2024GNN, Yang2024FDD, Shen2022, Huang2024Hetero, Wan2024Scalable}, but can help to significantly reduce the computational complexity of the ML models. More importantly, learning the phase guarantees inherent satisfaction of the analog beamforming constraint during the learning process without the need for any constant modular projection (which is non-differentiable), thereby maintaining the optimality of what has been learned and leading to better performance, as confirmed in the numerical results.
    \item Our graph accurately captures the OFDM structure using a fully-connected bipartite topology, where a single analog node connects to multiple subcarrier nodes. We further design an efficient message-passing mechanism that outperforms existing GNN-based approaches \cite{GNN_wireless,Wang2024GNN,Yang2024FDD} in both performance and generalization. Leveraging the permutation equivariance and scalability of GNNs, our model generalizes to a large number of subcarriers without retraining.
    \item We explore pure ML approaches by studying two different GNN update mechanisms: a node-update GNN and an edge-update GNN, achieved by placing the representation for the digital beamformer of each subcarrier at either a subcarrier node or an edge. These design choices provide insights into how representation placement affects the applicability of different GNN structures in hybrid beamforming. To the best of our knowledge, this is the first use of edge-update GNNs in hybrid beamforming. Building upon these designs, we directly extend the proposed pure ML models to multi-user scenarios through problem reformulation and graph structure expansion, without modifying the GNN updating rules.
    \item We further proposed a novel hybrid structure combining ML with traditional signal processing methods to develop a GNN structure in which the digital beamformers are updated using a closed-form singular-value decomposition result while learning the analog beamformer only. We also design an attention-based aggregation mechanism to overcome the reliance on a single learned analog node and improve the generalization of this GNN structure. Our hybrid design is unique and offers valuable insights into the performance of combining ML and traditional optimization methods.
    \item Finally, we perform extensive performance comparison and ablation study of the three proposed GNNs against multiple traditional optimization and existing ML-based solutions. The results show that the proposed message-passing mechanism in our GNNs outperforms all existing methods in spectral efficiency, generalization ability with respect to the number of users and subcarriers, beam squinting resiliency, and computational efficiency, even under imperfect CSI. These advantages are particularly prominent when the number of antennas increases.
\end{itemize}

\section{Single User System Model and Problem Formulation}\label{section2}
\subsection{System and Signal Models}

\begin{figure}[!t]
\centering
\includegraphics[width=3.0in]{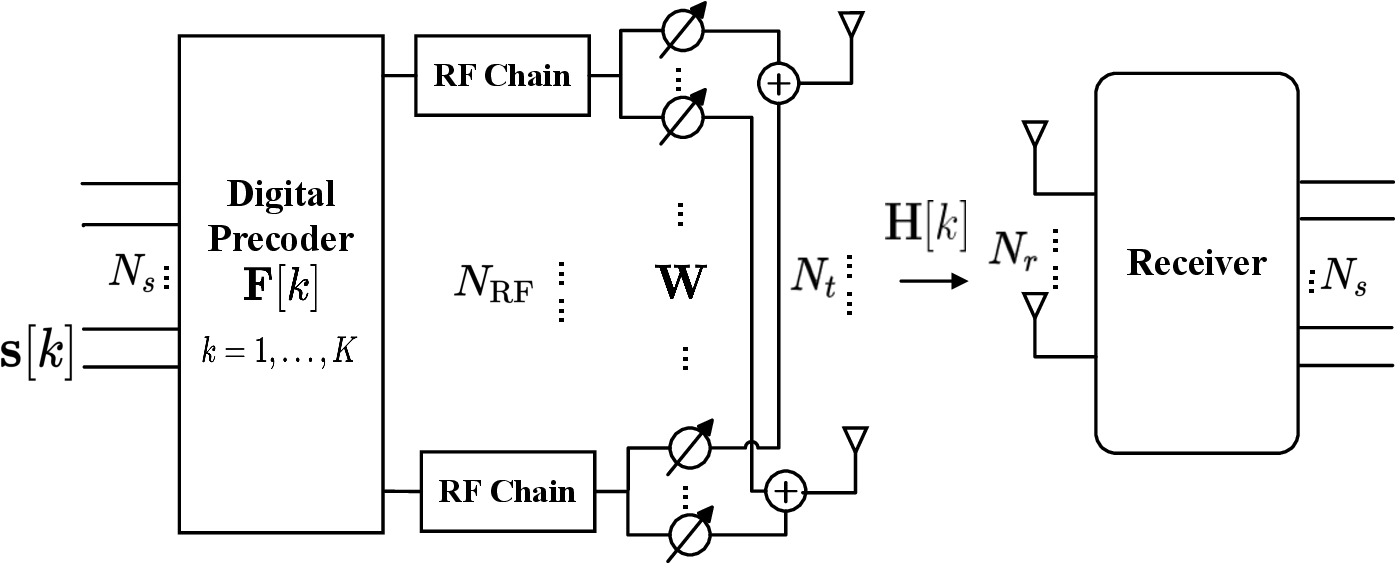}
\caption{Block diagram of a single-user MIMO-OFDM system with hybrid beamforming at the BS.}
\vspace{-7pt}
\label{fig_sys}
\end{figure}

We consider a single-user MIMO-OFDM system as shown in Fig. \ref{fig_sys}, where the base station (BS) equipped with $N_{\text{RF}}$ RF chains and $N_t$ antennas sends $N_s$ data streams to the user equipment (UE). The receiver has $N_r$ antennas, where $N_s \leq N_{\text{RF}} \leq N_r \ll N_t$ because of hardware constraint. The system employs OFDM where the transmission bandwidth $B$ is divided into $K$ subcarriers with equal widths, and the BS employs hybrid beamforming to transmit data to the UE.

The hybrid beamformer consists of a digital baseband beamformer $\mathbf{F}[k] \in \mathbb{C}^{N_{\text{RF}} \times N_s}$ for each subcarrier $k$, and an analog RF beamformer $\mathbf{W} \in \mathbb{C}^{N_t \times N_{\text{RF}}}$ shared among all subcarriers. The transmitted signal vector on the $k$-th subcarrier is
{\setlength{\abovedisplayskip}{4pt}
 \setlength{\belowdisplayskip}{4pt}
\begin{align}
    \mathbf{x}[k] = \sqrt{P_t}\mathbf{WF}[k]\mathbf{s}[k], \label{trans_sig}
\end{align}
}where $P_t$ is the averaged transmit power per subcarrier, and $\mathbf{s}[k]$ is the normalized $N_s \times 1$ symbol vector transmitted in each subcarrier $k = 1, 2, ..., K$, where $\mathbb{E}[\mathbf{s}[k]\mathbf{s}^*[k]]=\mathbf{I}_{N_s}$. {For the single-user setting, we assume equal power allocation across all subcarriers to focus on beam squinting mitigation, following \cite{ICD,AMO,sixmethods}. Simulation results under relaxed per-subcarrier power constraint with upper-bounded power budget are also provided for comparison in Section \ref{section9}.}

Accordingly, the transmit power constraint per subcarrier imposes a normalization on the beamformer matrices as 
{\setlength{\abovedisplayskip}{4pt}
 \setlength{\belowdisplayskip}{4pt}
\begin{align}
    ||\mathbf{WF}[k]||_F^2=1\label{power_constraint}.
\end{align}
}Furthermore, while the digital beamformer matrices $\mathbf{F}[k]$ can have complex-valued elements, the analog beamformer matrix W is restricted to having its elements with fixed magnitude because of phase array implementation. This condition leads to the constant modulo constraint as
{\setlength{\abovedisplayskip}{6pt}
 \setlength{\belowdisplayskip}{6pt}
\begin{align}
\left|\left[\mathbf{W}\right]_{i,j}\right|=1,\quad \forall i,j,\label{modulo_constraint}
\end{align}
}which can also be written as
{\setlength{\abovedisplayskip}{6pt}
 \setlength{\belowdisplayskip}{6pt}
\begin{align}
    \mathbf{W}=e^{j\mathbf{\Phi}},\label{modulo_constraint2}
\end{align}
}where $\mathbf{\Phi} \in \mathbb{R}^{N_t \times N_{\text{RF}}}$ is the phase matrix, with unwrapped phase elements $-\infty \leq \mathbf{\Phi}_{i,j} \leq \infty$.

The received signal on the $k$-th subcarrier is
{\setlength{\abovedisplayskip}{4pt}
 \setlength{\belowdisplayskip}{4pt}
\begin{align}
    \mathbf{y}[k] = \sqrt{P_t}\mathbf{H}[k]\mathbf{WF}[k]\mathbf{s}[k]+\mathbf{n}[k],
\end{align}
}where $\mathbf{H}[k]$ is the channel matrix on the $k$-th subcarrier, and $\mathbf{n}[k]\sim\mathcal{CN}(0,\sigma_n^2\mathbf{I}_{N_r})$ is the additive white Gaussian noise.

\subsection{Channel Model}
We adopted a wideband clustered double-directional channel model \cite{channel}:
{\setlength{\abovedisplayskip}{4pt}
 \setlength{\belowdisplayskip}{5pt}
\begin{align}
    \mathbf{H}[k] = \frac{1}{\sqrt{N_{\text{cl}}N_{\text{ray}}}} 
    \sum_{i=1}^{N_{\text{cl}}} \sum_{l=1}^{N_{\text{ray}}} 
    \alpha_{il,k}\beta_{il,k} \mathbf{a}_{k}^r(\phi_{il}^r, \theta_{il}^r) 
    \mathbf{a}_{k}^t(\phi_{il}^t, \theta_{il}^t)^*\label{channel},
\end{align}
}where $N_{\text{cl}}$ and $N_{\text{ray}}$ represent the number of clusters and rays within each cluster. $\alpha_{il,k}$ denotes the complex path gain of the $l$-th ray in the $i$-th cluster, and $\beta_{il,k}=e^{-j2\pi\tau_{il}f_k}$ represents the propagation path delay component. The angles $(\phi^r_{il}, \theta^r_{il})$ and $(\phi^t_{il}, \theta^t_{il})$ represent the azimuth and elevation angles of arrival and departure, respectively. Considering uniform planar arrays (UPA), the array response vector corresponding to the $l$-th ray in the $i$-th cluster is
{\setlength{\abovedisplayskip}{5pt}
 \setlength{\belowdisplayskip}{4pt}
\begin{align}
    \mathbf{a}_k(\phi_{il}, \theta_{il}) = &\left[ 1, ..., e^{j\frac{2\pi}{\lambda_k}d\left(p\sin\phi_{il}\sin\theta_{il}+q\cos\theta_{il}\right)},...,\right.\nonumber\\
    &\left.e^{j\frac{2\pi}{\lambda_k}d\left(\left(M-1\right)\sin\phi_{il}\sin\theta_{il}+\left(N-1\right)\cos\theta_{il}\right)}\right]^T,
\end{align}
}where $d$ and $\lambda_k$ are the antenna spacing and the signal wavelength, $0 \leq p < N$ and $0 \leq q < M$ are the antenna indices in the 2D plane. 

\subsection{Beam Squinting Effect}
\begin{figure}[!t]
    \centering
    \subfloat[]{%
        \hspace{-0.04\linewidth}\includegraphics[width=0.45\linewidth]{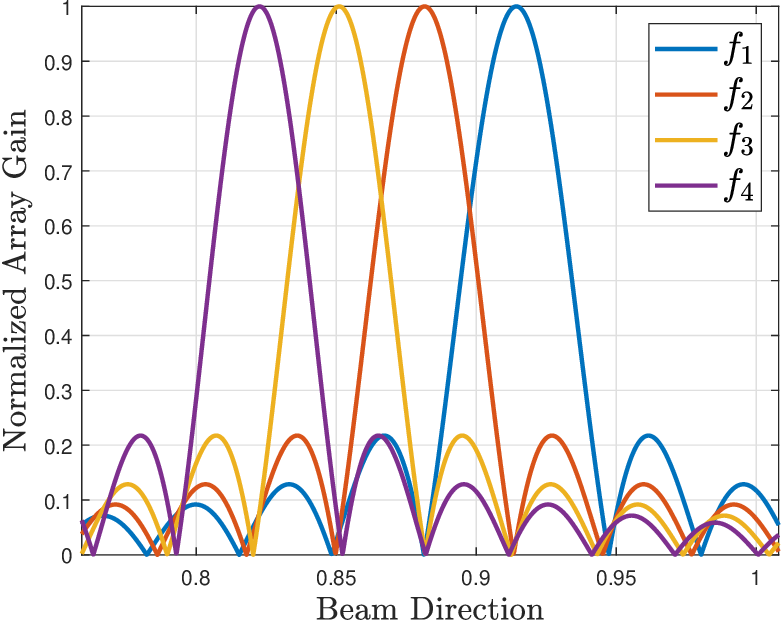}%
        \label{fig:beam direction}%
    }
    \subfloat[]{%
        \hspace{0.04\linewidth}
        \includegraphics[width=0.45\linewidth]{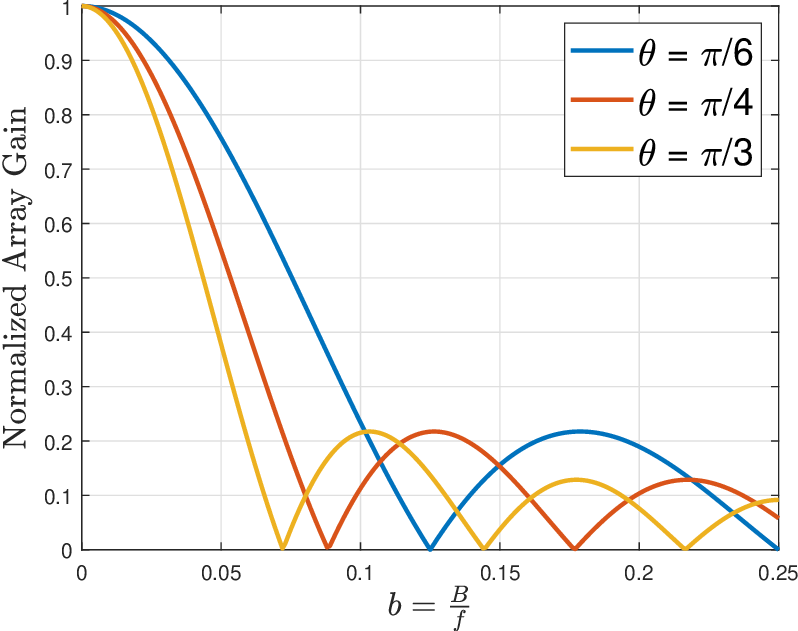}%
        \label{fig:array gain}%
    }
    \caption{Beam squinting effect in a wideband system with the central frequency $f_c = 142\,\mathrm{GHz}$, bandwidth $B = 20\,\mathrm{GHz}$. (a) The beam's direction shifts across 4 subcarriers. (b) The normalized array gain versus the fractional bandwidth $b = B/f_c$ for angle of arrivals (AoAs) $\theta = \{\pi/6, \pi/4, \pi/3\}$. Typical narrowband settings \cite{Li2025, Huang2024Hetero} and the considered system are also marked in (b) for comparison.}
    \vspace{-13pt}
    \label{fig:beamsquint}
\end{figure}

{Beam squinting refers to the frequency-dependent array gain in wideband systems where the beam's direction shifts across subcarriers, leading to array gain degradation \cite{beamsquint}. This effect becomes more noticeable as the system bandwidth expands, as shown in Fig. \ref{fig:beamsquint}.} In hybrid beamforming systems, although the digital beamformer offers flexibility, the analog beamformer is constrained by a single set of phase shifters that is shared across all subcarriers, making it the cause for beam squinting. As a result, accurately characterizing beam squinting requires an explicit wideband modeling framework that accounts for the frequency-dependent behavior across subcarriers.

The shared analog beamformer is captured in the problem formulation that jointly optimizes the system across all subcarriers by enforcing a shared analog beamforming matrix $\mathbf{W}$ while allowing independent digital beamforming matrices $\mathbf{F}[k]$ for each subcarrier. The effectiveness of beam squinting mitigation depends on the algorithm to solve the problem and will serve as a metric to evaluate the algorithms' performance.

\subsection{Problem Formulation}
We focus on the design of the transmit beamformers at the BS, assuming the perfect channel state information (CSI). The achievable spectral efficiency can be expressed as
{\setlength{\abovedisplayskip}{4pt}
 \setlength{\belowdisplayskip}{4pt}
\begin{align}
    R = \frac{1}{K}\sum_{k=1}^K \log_2\biggl[\det&\biggl(\mathbf{I}_{N_r} + \frac{P_t}{\sigma_n^2}\mathbf{H}[k]\mathbf{WF}[k] \biggr.\biggr.\nonumber \\
    &\quad \biggl.\biggl. \times \mathbf{F^*}[k]\mathbf{W^*H^*}[k]\biggr)\biggr] (\text{bps/Hz}) \label{DataRate}.
\end{align}
}Then the beamforming design problem can be posed as
{\setlength{\abovedisplayskip}{4pt}
 \setlength{\belowdisplayskip}{4pt}
\begin{align}
    \max_{\mathbf{W, F}[k]} \ R \quad\quad
	\text{s.t.} \ (\ref{power_constraint}), (\ref{modulo_constraint}).\label{prob}
\end{align}
}

This hybrid beamforming problem is non-convex due to the constant modulo constraint in (\ref{modulo_constraint}). Traditional signal processing methods typically address this non-convexity through alternating optimization. In each algorithm step, the analog beamformer is reconstructed by extracting only the phase matrix of the resulting complex-valued matrix in each algorithm step to satisfy the constant modulo constraint \cite{sixmethods, AMO, ICD}, which can result in suboptimality. In our approach, we choose to treat the phases of the analog beamformer as the unknown variables directly. As such, the problem formulation becomes
{\setlength{\abovedisplayskip}{4pt}
 \setlength{\belowdisplayskip}{4pt}
\begin{align}
\hspace{-0.1cm}
    \max_{\mathbf{\Phi, F}[k]} \  &R(\mathbf{\Phi},\mathbf{F}[k]) \triangleq\frac{1}{K}\sum_{k=1}^K \log_2\left[\det\left(\mathbf{I}_{N_r} + \frac{P_t}{\sigma_n^2}\mathbf{H}[k]\mathbf{e^{j\mathbf{\Phi}} F}[k]\right.\right.\nonumber\\
    &\left.\left.\quad\quad\quad\quad\quad\quad\times \mathbf{F^*}[k]\left(e^{j\mathbf{\Phi}}\right)^*\mathbf{H^*}[k]\right)\right]\nonumber\\
	\text{s.t.} \  
			&\left|\left|e^{j\mathbf{\Phi}}\mathbf{F}[k]\right|\right|_F^2=1.\label{prob_psi}
\end{align}}Formulation (\ref{prob_psi}) is equivalent to formulation (\ref{prob}) but has fewer constraints because of the change of variables.

\begin{figure}[!t]
\centering
\includegraphics[width=2.2in]{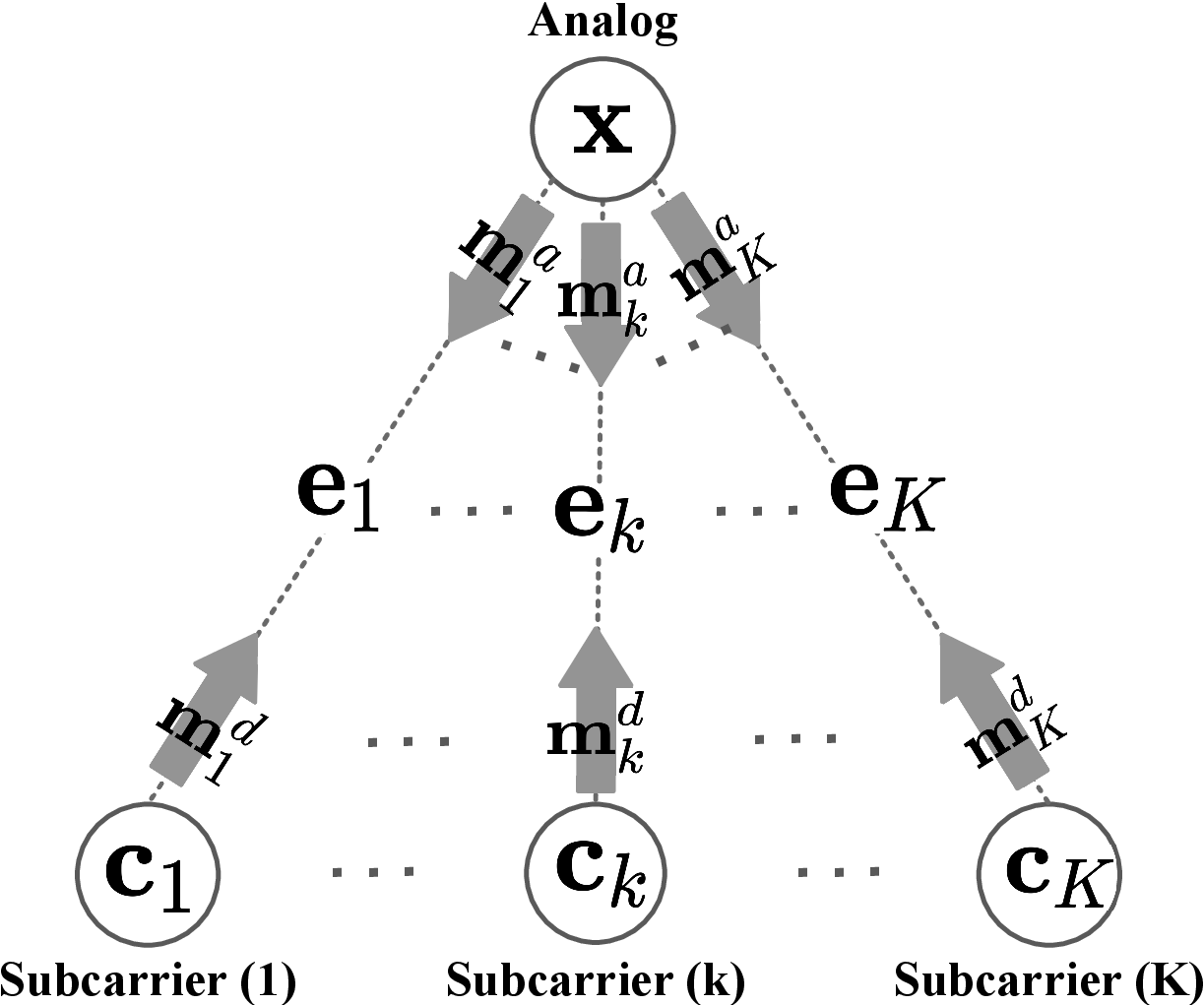}
\caption{Bipartite graph model for a hybrid beamforming structure with two types of nodes: analog and subcarrier nodes. Information embedded at the analog node is represented as vector $\mathbf{x}$, at each subcarrier node as $\mathbf{c}_k$, and on each edge as $\mathbf{e}_k$. Each subcarrier node and edge corresponds to a subcarrier $k$, while the analog node is shared among all subcarriers. The proposed GNNs based on this graph model employ a message-passing mechanism where messages are denoted as $\mathbf{m}_{k}^a$ and $\mathbf{m}_{k}^d$.}
\vspace{-7pt}
\label{fig:graph_model}
\end{figure}

While traditional optimization methods exploit the problem's mathematical structure, they require re-optimization for every new CSI, which is computationally intensive. In contrast, ML methods are not constrained by problem structure and only rely on dataset statistics. Once trained, they only require forward computations for any new, unseen CSI, which makes the computational process much simpler and faster. As such, we will examine the use of ML to tackle this problem.

\section{Message-passing GNN structures}\label{section3}

We propose graph-based learning structures for hybrid beamforming design in wideband systems. We begin by motivating our choice of GNNs over other ML architectures. We then construct a graph model that serves as the foundation for the GNN structures introduced in the following sections. Finally, we describe the training process applicable to all proposed GNN structures.

\subsection{Motivations of Selecting GNNs}
Traditional fully-connected neural networks (FNNs) require fixed-length inputs whose dimensions scale with system size, such as the number of subcarriers, and must be retrained for different configurations, therefore allowing little to no generalization or scaling ability. Convolutional neural networks (CNNs) are designed for grid-structured data and have similar issues with varying input sizes. Techniques like padding or interpolation can be used to align input dimensions, but often degrade performance.

In contrast, GNNs utilize the system’s underlying structural graph, where nodes represent system elements such as subcarriers in our problem. Reordering nodes does not affect the underlying graph mapping, and each node type typically shares a common update function, so the number of trainable parameters does not depend on the graph size \cite{GNN_wireless}. This scalability allows GNNs to generalize across systems with varying subcarrier configurations, which is a powerful and desirable property. Guided by these insights, we choose GNN as the architecture of choice for designing hybrid beamforming for an OFDM system. The underlying graph model in a GNN can capture the relationship between analog and digital beamformers in a multicarrier OFDM structure while preserving its inherent symmetry, allowing strong scalability.

\begin{figure}[!t]
\centering
\includegraphics[width=3.5in]{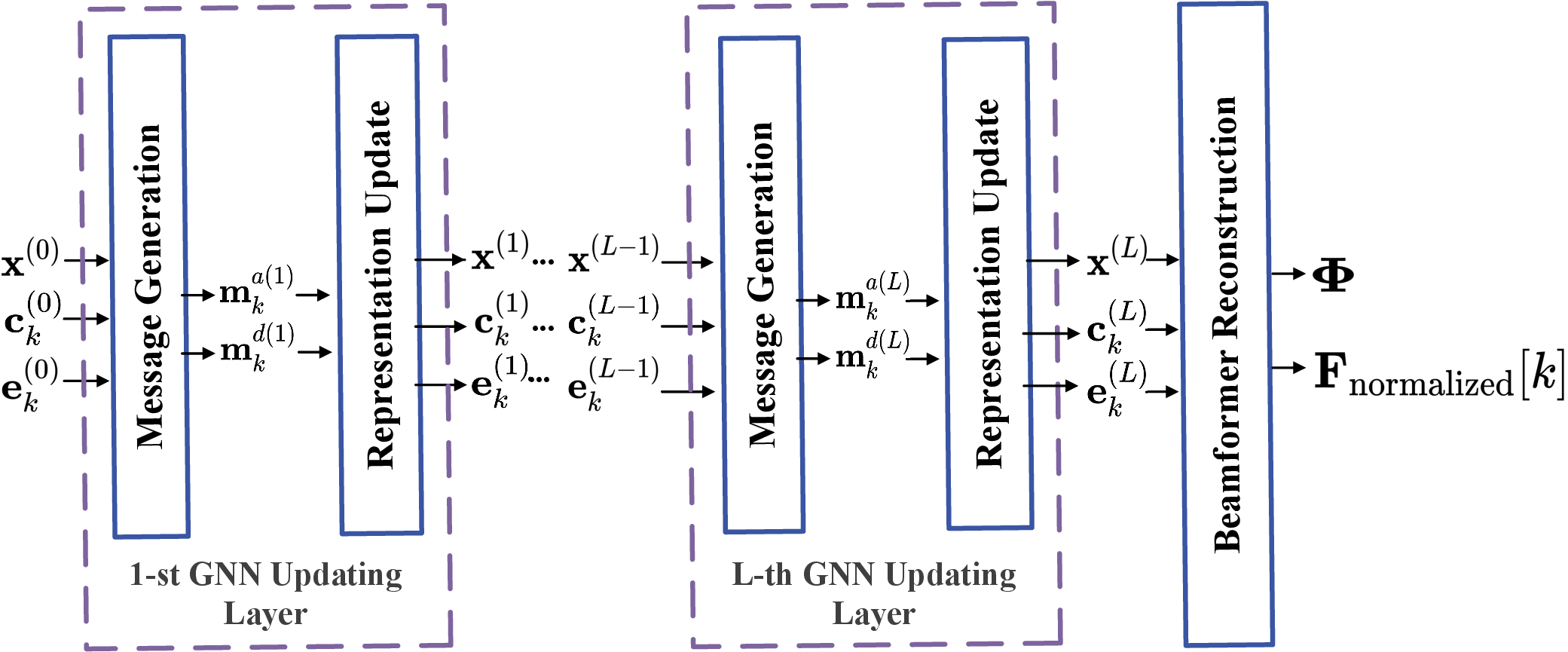}
\caption{Layer-wise GNN architecture overview. Each updating layer consists of a message generation and a representation update step. The final layer is a beamformer reconstruction to produce the output. The inputs ($\mathbf{x}^{(0)},\mathbf{c}_k^{(0)},\mathbf{e}_k^{(0)}$) depend on the specific proposed GNN structures. The specific configurations in each updating layer for different proposed GNN structures are detailed in Figs. \ref{fig:NU_stru}--\ref{fig:AN_stru}.}
\vspace{-7pt}
\label{fig:GNN_stru}
\end{figure}

\subsection{Message-passing Graph Structure}
To better understand the generalization property of a GNN, we examine the subcarrier-level behavior and establish the following permutation equivariant properties. We first show that the original problem exhibits this property, and when mapped to a GNN, the GNN also maintains this property.

\textit{Proposition 1:} The formulated problem in (10) is permutation equivariant with respect to the subcarriers, RF chains, and data streams.

\textit{Proof:} See Appendix~A. \hfill $\blacksquare$

Proposition 1 implies that permuting the subcarrier, RF chain, and data stream indices does not affect the optimal solution of the problem. {Since this work focuses on addressing beam squinting effects rather than exhaustively exploiting all possible equivariance properties, we mainly leverage the subcarrier-level equivariance to construct the proposed graph model.}

We employ a bipartite, undirected graph, where a single ``analog node" connects to $K$ ``subcarrier nodes", each representing a subcarrier. As shown in Fig. \ref{fig:graph_model}, we embed information at the analog node as $\mathbf{x} \in \mathbb{R}^{N_tN_{\text{RF}}}$, at each subcarrier node $k$ as $\mathbf{c}_k \in \mathbb{R}^{2N_{\text{RF}}N_s}$, and on each edge as $\mathbf{e}_k \in \mathbb{R}^{2N_tN_r}$. Note that not all of these embeddings, also called representations, need to be present in a GNN structure, and we will propose three different GNN structures that use different subsets of these representations. The specific design of each proposed GNN structure is illustrated in Figs. \ref{fig:NU_stru}--\ref{fig:AN_stru} and described in Sections IV and V. Before introducing these structures in detail, we first describe the general form of message-passing used in our framework.

Figure \ref{fig:GNN_stru} provides an overview of the layer-wise GNN architecture, which stacks $L$ updating layers for progressively learning and refining representations. Each layer employs a message-passing mechanism, where nodes exchange messages with their neighbors to update their representations. The design of these mechanisms for message passing and representation updating will define the learning process in the GNN and is important for the final performance and generalization ability. Specifically, we generate messages $\mathbf{m}^a_{k} \in \mathbb{R}^{N_tN_{\text{RF}}}$ to carry information from the analog node $\mathbf{x}$ to the $k$-th subcarrier node $\mathbf{c}_k$, while $\mathbf{m}^d_{k} \in \mathbb{R}^{2N_{\text{RF}}N_s}$ transmit information in the reverse direction. At a high level, we can summarize the updating operations in our proposed GNNs as:
{\setlength{\abovedisplayskip}{4pt}
 \setlength{\belowdisplayskip}{4pt}
\begin{align}
    \mathbf{b}_k^{(l)} = f_{\mathbf{b}}\left(\mathbf{a}_k^{(l-1)}, \phi(\cdot)\right), \quad  \forall~\mathbf{a}_k, \mathbf{b}_k \in \mathcal{A},  \; \forall k, \label{GNN_mapping}
\end{align}}where $\mathcal{A} = \{\mathbf{x}, \mathbf{e}, \mathbf{c}, \mathbf{m}^a, \mathbf{m}^d\}$ is a set of all representations and messages. Function $\phi(\cdot)$ denotes a permutation invariant aggregation function that is consistently applied to the selected node's neighbors, thus preserving the permutation equivariance of the overall updating operation. Each function $f_{\mathbf{b}}$ is typically realized by a multi-layer perceptron (MLP) network and is shared across all subcarriers $k$. The specific design for the mapping in (\ref{GNN_mapping}) varies with each GNN structure and will be discussed in Section \ref{section5} and \ref{section6}.

Based on (\ref{GNN_mapping}), all our designed GNN structures exhibit the following property:

\textit{Proposition 2:} The output of each GNN updating layer is permutation equivariant with respect to the subcarrier order. As such, the final outputs of each GNN, including the analog and digital beamformers, are permutation equivariant.

\textit{Proof:} See Appendix~B. \hfill $\blacksquare$

Proposition 2 highlights the generalization ability of the proposed GNNs across subcarrier-level variations, enabling adaptation to dynamic system configurations.


\subsection{Considerations in Designing GNN Structures}\label{sec: considerations}

The graph structure in Fig. \ref{fig:graph_model} captures the underlying communication system by representing the analog beamformer as a single node $\mathbf{x}$, shared across all subcarriers, and therefore is learned via the node representation at this analog node. The main consideration in designing a specific GNN model lies in how to represent the digital beamformers. We explore three strategies for representing the digital beamformers: (i) using the digital beamformer node representations (node-update structure), (ii) using edge representations (edge-update structure), and (iii) a hybrid structure which only learns the analog beamformer and bypasses learning the digital beamformer entirely by applying the traditional SVD-based digital beamforming solutions derived from the learned analog beamformer.

Different GNNs are expected to show varying performance and complexities. The first two GNN structures will illustrate the differences in node or edge updates, and their impact on the overall performance. The third GNN structure will examine the difference between learning everything vs. using known traditional optimization results in some steps. Extensive analysis and ablation study of these structures will offer valuable insights for designing GNN structures in practical wireless systems, with impacts on achievable performance, computational complexity, and scalability.

\subsection{Loss Function and Unsupervised Training}
For each GNN structure, we train the model such that the GNN can effectively update the representations to achieve a high average data rate, as formulated in (\ref{prob_psi}). Let $\mathbf{\Omega}$ denote all trainable parameters in a GNN for message generation and representation update. During offline training, we train the GNN in an unsupervised manner to optimize $\mathbf{\Omega}$ by minimizing the loss function derived from (\ref{prob_psi}) as:
{\setlength{\abovedisplayskip}{4pt}
 \setlength{\belowdisplayskip}{4pt}
\begin{align}
\text{Loss}(\mathbf{\Omega}) =&-R(\mathbf{\Phi},\mathbf{F}_{\text{normalized}}[k]).
\label{loss}
\end{align}}

This loss function is computed using the GNN's outputs, $\mathbf{\Phi}$ and $\mathbf{F}_{\text{normalized}}[k]$, along with the available CSI. To ensure that the outputs satisfy the constraints specified in (\ref{prob_psi}), we apply normalization steps during the beamformer reconstruction process to meet the required power constraint.

We then minimize the loss in (\ref{loss}) using a mini-batch stochastic gradient descent (SGD) approach as:
{\setlength{\abovedisplayskip}{4pt}
 \setlength{\belowdisplayskip}{4pt}
\begin{equation}
\mathbf{\Omega}^{(i+1)} \leftarrow \mathbf{\Omega}^{(i)} - \eta^{(i)}\nabla_{\mathbf{\Omega}}\mathbb{E}_{\mathcal{B}}\left[\text{Loss}(\mathbf{\Omega})\right]\label{minibatch},
\end{equation}}where $\eta$ is the learning rate, and $\mathcal{B}$ denotes the mini-batch set.

Because of the permutation equivariance and scalability of GNNs, it suffices to train on a small number of subcarriers, given sufficient CSI samples spanning a wide frequency range and channel conditions. Then during the online inference, the number of subcarrier nodes can vary as needed to suit different OFDM system configurations.

\section{Node and Edge Update GNN Structures}\label{section5}
In this section, we consider using pure machine learning to design hybrid beamformers. We propose two GNN structures: node-update GNN (NU-GNN) and edge-update GNN (EU-GNN). Both approaches represent the analog beamformer at the analog node but differ in handling the digital beamformers.

NU-GNN embeds the digital beamformers in the subcarrier nodes by using $\{\mathbf{c}_1,...,\mathbf{c}_K\}$ as the digital beamformer representations, while keeping the edge features fixed. The edge features are the vectorized CSI, such that
{\setlength{\abovedisplayskip}{2pt}
 \setlength{\belowdisplayskip}{4pt}
\begin{align}
\hspace{-0.31cm}
\mathbf{e}_k=\mathbf{h}_k=\left[ \text{vec} \left( \text{Re} \left\{ \rho\mathbf{H}[k] \right\} \right)^T, \text{vec} \left( \text{Im} \left\{ \rho\mathbf{H}[k] \right\} \right)^T \right]^T,\label{edge}
\end{align}}where $\rho=\sqrt{\frac{P_t}{\sigma_n^2}}$ scales the channels to mitigate the small magnitudes caused by pathloss. This ensures numerically suitable inputs for training without impacting the final sum-rate, as the scaling is compensated during computation.

In contrast, EU-GNN embeds the digital beamformers in the edge representations $\{\mathbf{e}_1,...,\mathbf{e}_K\}$, without using any subcarrier node representations. We use the vectorized CSI in (\ref{edge}) as the initial input of the edge updating process:
{\setlength{\abovedisplayskip}{4pt}
 \setlength{\belowdisplayskip}{2pt}
\begin{align}
    \mathbf{e}^{(0)}_k=\mathbf{h}_k,\quad \forall k=1,...,K.
\end{align}}

By exploring both structures, we aim to understand how the update mechanism affects system performance and the tradeoff between learning flexibility and computational complexity.


\subsection{Node Update GNN (NU-GNN) Structure}\label{sec:NUGNN}

\begin{figure}[!t]
\centering
\includegraphics[width=2.7in]{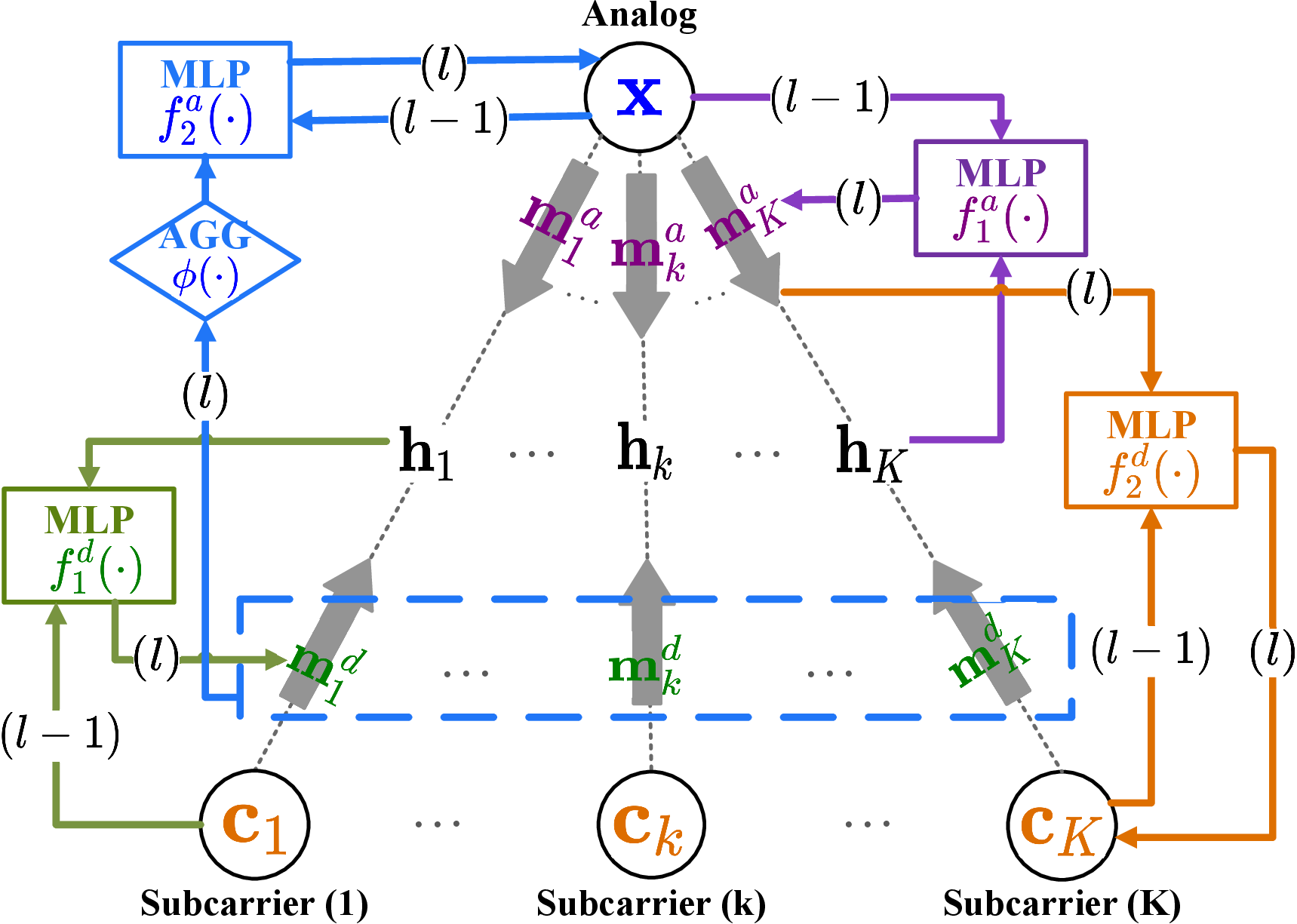}
\caption{NU-GNN Structure: During the message-passing process, two MLPs $f_1^a(\cdot)$ and $f_1^d(\cdot)$ generate messages $\mathbf{m}_{k}^a$ and $\mathbf{m}_{k}^d$, which are marked in purple and green. Then, two other MLPs $f_2^a(\cdot)$ and $f_2^d(\cdot)$ update the node representations $\mathbf{x}$ and $\mathbf{c}_k$, marked in blue and orange. The black colored $\mathbf{h}_k$ represents the edge feature, which does not get updated in this structure.}
\vspace{-7pt}
\label{fig:NU_stru}
\end{figure}

\begin{figure}[!t]
\centering
\includegraphics[width=2.7in]{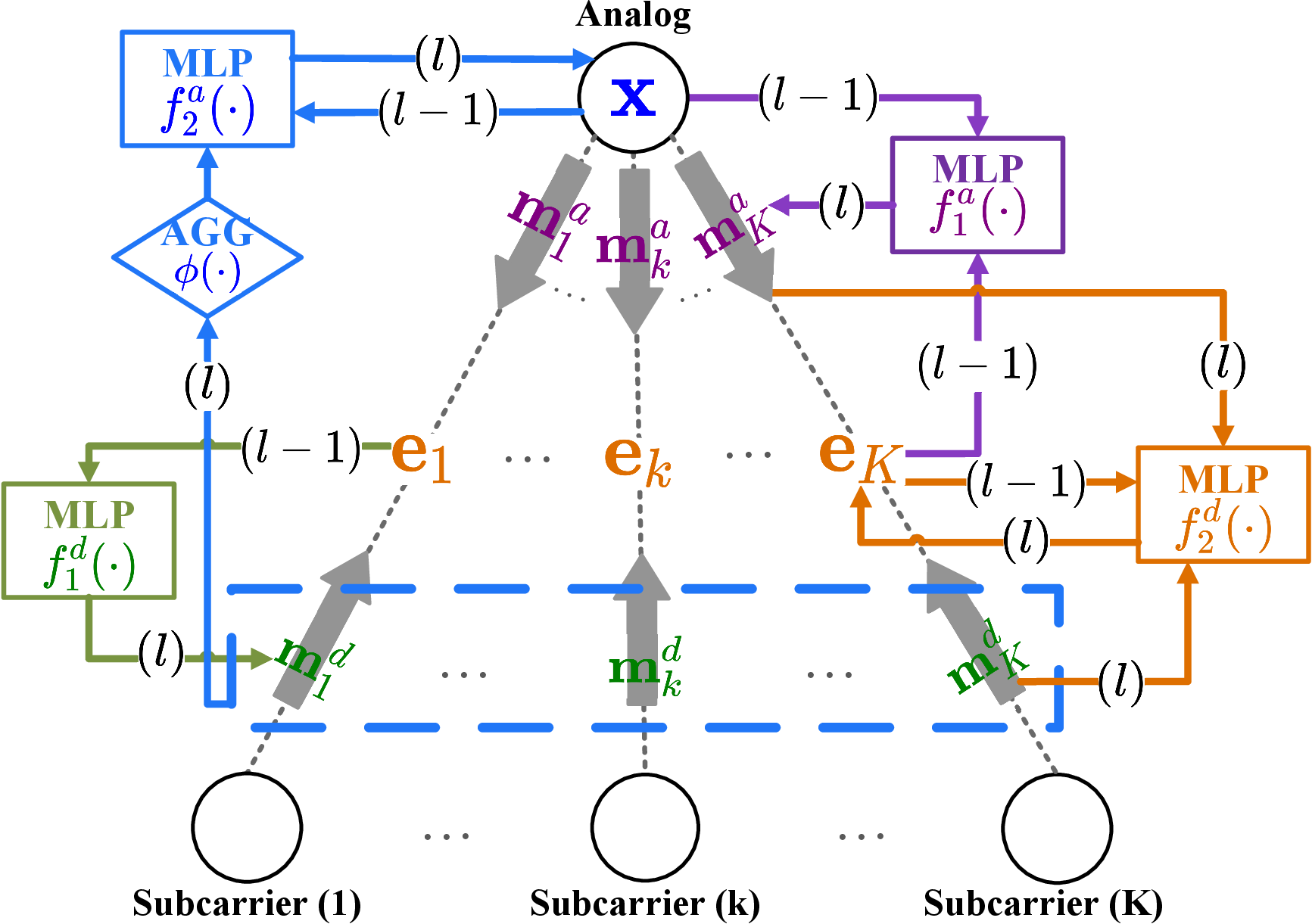}
\caption{EU-GNN Structure: During the message-passing process, two MLPs $f_1^a(\cdot)$ and $f_1^d(\cdot)$ generate messages $\mathbf{m}_{k}^a$ and $\mathbf{m}_{k}^d$, marked in purple and green. We also used two other MLPs $f_2^a(\cdot)$ and $f_2^d(\cdot)$ to update the node representation $\mathbf{x}$ and the edge representation $\mathbf{e}_k$, which are marked in blue and orange. In this structure, the CSI $\mathbf{h}_k$ is only used as the initial value of the edge representation, which will be updated during the learning process by MLP $f_2^d(\cdot)$ marked in orange.}
\vspace{-7pt}
\label{fig:EU_stru}
\end{figure}

Next, we design the message generation and representation update methods. These designs are important as they directly affect the GNN performance and computational complexity.

At the $l$-th updating layer of the GNN, the messages sent from the analog node to subcarrier node $k$, and from subcarrier node $k$ to the analog node, can be generated as
\setlength{\abovedisplayskip}{4pt}
\setlength{\belowdisplayskip}{4pt}
\begin{align}
    &\hspace{-0.31cm}
    \begin{aligned}\mathbf{m}_{k}^{a(l)}=f_1^a(\mathbf{h}_k,\mathbf{x}^{(l-1)})\label{NU_ma}
    \end{aligned} \\
    &\hspace{-0.31cm}
    \begin{aligned}\mathbf{m}_{k}^{d(l)}=f_1^d(\mathbf{h}_k,\mathbf{c}_k^{(l-1)}),\label{NU_md}
    \end{aligned}
\end{align}
where $\textit{f}_1^{\,a}(\cdot)$ and $\textit{f}_1^{\,d}(\cdot)$ are two MLPs, as shown in Fig. \ref{fig:NU_stru} in purple and green, to generate the messages at the analog node and subcarrier nodes, respectively.

After generating the messages, each node receives them via the connecting edges, aggregates these incoming messages from its neighbors, and then updates its representation vector accordingly as follows.
\setlength{\abovedisplayskip}{4pt}
\setlength{\belowdisplayskip}{4pt}
\begin{align}
&\mathbf{x}^{(l)}=f_2^a(\mathbf{x}^{(l-1)}, \phi({\mathbf{m}_{k}^{d(l)}})_{k\in\mathcal{N}(\mathbf{x})})\\
&\mathbf{c}_k^{(l)}=f_2^d(\mathbf{c}_k^{(l-1)}, \mathbf{m}_{k}^{a(l)})
\end{align}
$\textit{f}_2^{\,a}(\cdot)$ and $\textit{f}_2^{\,d}(\cdot)$ are two other MLPs as shown in Fig. \ref{fig:NU_stru}, marked in blue and orange, to update the analog and digital beamformer representations, respectively.
$\mathcal N(\mathbf{x})$ represents the set of neighboring nodes of $\mathbf{x}$, and $\phi(\cdot)$ is the element-wise mean function used to combine the information coming into the same node. While both mean and max aggregation functions are permutation invariant, mean aggregation captures the collective impact of beam squinting by incorporating all subcarriers rather than the most dominant one.

\subsection{Edge Update GNN (EU-GNN) Structure}

In the EU-GNN, we design the message generation and representation update as follows.

At the $l$-th GNN layer, the messages are generated as
\setlength{\abovedisplayskip}{4pt}
\setlength{\belowdisplayskip}{4pt}
\begin{align}
&\mathbf{m}_{k}^{a(l)}=f_1^a(\mathbf{e}_k^{(l-1)},\mathbf{x}^{(l-1)})\label{EU:ma}\\
&\mathbf{m}_{k}^{d(l)}=f_1^d(\mathbf{e}_k^{(l-1)})\label{EU:md}
\end{align}
where $\textit{f}_1^{\,a}(\cdot)$ and $\textit{f}_1^{\,d}(\cdot)$ are two MLPs as shown in Fig. \ref{fig:EU_stru} marked in purple and green, respectively.


Similar to NU-GNN, we employ the generated messages as the inputs to the representation updates as follows.
\setlength{\abovedisplayskip}{4pt}
\setlength{\belowdisplayskip}{4pt}
\begin{align}
&\mathbf{x}^{(l)}=f_2^a(\mathbf{x}^{(l-1)}, \phi({\mathbf{m}_{k}^{d(l)}})_{k\in\mathcal{N}(\mathbf{x})})\\
&\mathbf{e}_k^{(l)}=f_2^d(\mathbf{e}_k^{(l-1)}, \mathbf{m}_{k}^{a(l)}, \mathbf{m}_{k}^{d(l)})\label{EU:e}
\end{align}
Here $\textit{f}_2^{\,a}(\cdot)$ and $\textit{f}_2^{\,d}(\cdot)$ are two MLPs marked in blue and orange, respectively, as shown in Fig. \ref{fig:EU_stru}. The node representation $\mathbf{x}$ is updated by aggregating the messages from its neighbor nodes, and the edge representation $\mathbf{e}_k$ is updated by aggregating the messages flowing into this edge.

\subsection{Beamformer Reconstruction}
To obtain the analog beamforming matrix, the final layer's analog node representation $\mathbf{x}^{(L)}$ is reshaped into the phase matrix $\mathbf{\Phi}$:
{
\setlength{\abovedisplayskip}{-10pt}
\setlength{\belowdisplayskip}{-10pt}
\begin{equation}
\mathbf{\Phi}=\text{reshape}(\mathbf{x}^{(L)},(N_t,N_{\text{RF}})).\label{analog_reconstruct}
\end{equation}
}By directly learning the phase of the analog beamformer, the GNN output in (\ref{analog_reconstruct}) inherently satisfies the constant modulus constraint (\ref{modulo_constraint}) without requiring any further projection or approximation, which would harm the learning performance.

To satisfy the power constraint in (\ref{prob_psi}), the updated node representations $\mathbf{c}_k$, or the edge representations $\mathbf{e}_k$, from the final layer $L$ of the GNNs must be further processed. The digital representations $\mathbf{c}_k^{(L)}$ (or $\mathbf{e}_k^{(L)}$) are reassembled into a complex-valued matrix $\mathbf{F}_k$ and normalized to fulfill the power constraint in (\ref{prob_psi}):
{\setlength{\abovedisplayskip}{4pt}
\setlength{\belowdisplayskip}{4pt}
\begin{equation}
\mathbf{F}_{\text{normalized}}[k] = \frac{\mathbf{F}_k}{|| \mathbf{F}_k e^{j\mathbf{\Phi}}||_F^2},\label{digital_reconstruct}
\end{equation}}where, $\mathbf{F}_k=\text{reshape}\left(\mathbf{c}_k^{(L)}\left[1:N_{\text{RF}}\times N_s\right],\left(N_{\text{RF}},N_s\right)\right) + j\ \text{reshape}\left(\mathbf{c}_k^{(L)}\left[N_{\text{RF}}\times N_s:2N_{\text{RF}}\times N_s\right],\left(N_{\text{RF}},N_s\right)\right)$ in the NU-GNN. In the EU-GNN, $\mathbf{c}_k^{(L)}$ can be replaced by the edge representation $\mathbf{e}_k^{(L)}$ in the same reconstruction process.

\section{Analog Node GNN with Attention Structure}\label{section6}
\begin{figure}[!t]
\centering
\includegraphics[width=2.7in]{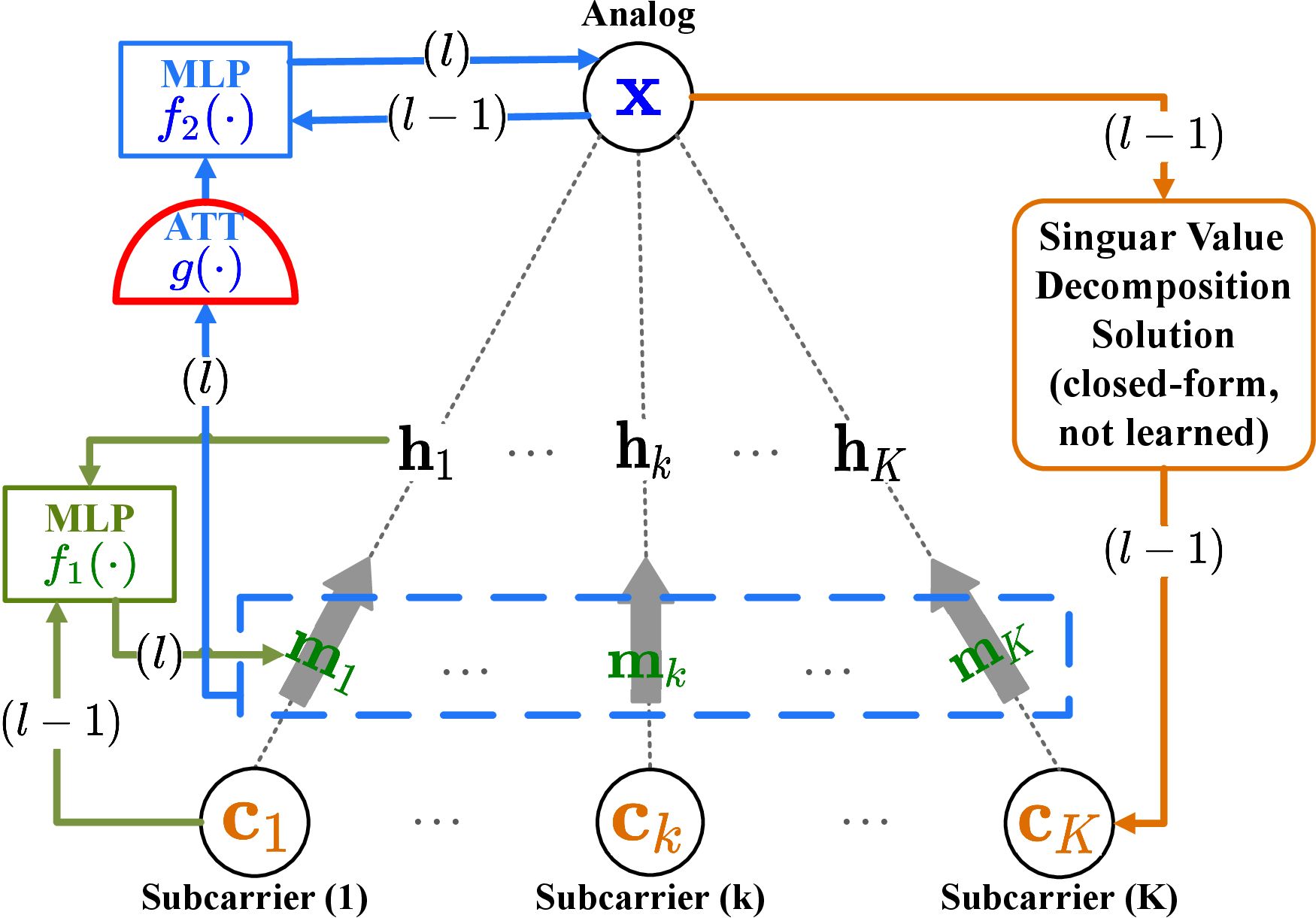}
\caption{AN-GNN Structure: Before updating the messages, we use the singular-value decomposition solution in (\ref{waterfilling_digital2}) and (\ref{node_feature}) to generate the digital beamformer representations $\mathbf{c}_k^{(l-1)}$ at the subcarrier nodes. Two MLPs $f_1(\cdot)$ and $f_2(\cdot)$ are deployed to generate the messages $\mathbf{m}_{k}$ and to update the analog node representation $\mathbf{x}$, marked in green and blue, respectively. An attention aggregation $g(\cdot)$ outlined in red is applied to aggregate the gathered information across all subcarriers.}
\vspace{-7pt}
\label{fig:AN_stru}
\end{figure}
Unlike the previous two GNN structures, which use pure ML to jointly learn both analog and digital beamformers, here we design a different structure termed analog-GNN (AN-GNN), which combines machine learning with traditional optimization. Specifically, we update analog node representation $\mathbf{x}$ via learning, derive digital beamformer representations $\mathbf{c}_k$ via closed-form expressions, and fix edge features $\mathbf{e}_k=\mathbf{h}_k$ as in (\ref{edge}). Next, we reformulate problem (\ref{prob_psi}) to solve for the digital beamformers and detail the AN-GNN's learning mechanisms.


\subsection{Problem Reformulation}

\textit{1) Digital Beamformer Solution:} For a given analog beamformer, the digital beamformer in (\ref{prob_psi}) can be solved in closed-form by constructing an effective channel $\mathbf{H}_{\text{eff}}[k]=\mathbf{H}[k]e^{j\mathbf{\Phi}}\left(\left(e^{j\mathbf{\Phi}}\right)^*e^{j\mathbf{\Phi}}\right)^{-\frac{1}{2}}$. Let $\tilde{\mathbf{F}}[k]=\left(\left(e^{j\mathbf{\Phi}}\right)^*e^{j\mathbf{\Phi}}\right)^{\frac{1}{2}}\mathbf{F}[k]$, the original problem in (\ref{prob_psi}) is
\begin{align}
    \max_{\mathbf{F}[k]} \  & \frac{1}{K}\sum_{k=1}^K \log_2[\det(\mathbf{I}_{N_r} + \frac{P_t}{\sigma_n^2}(\mathbf{H}_{\text{eff}}[k]\mathbf{\tilde{F}}[k]\times \mathbf{\tilde{F}^*}[k]\mathbf{H}^*_{\text{eff}}[k]))]\nonumber\\
	\text{s.t.} \  & ||\mathbf{\tilde{F}}[k]||_F^2=1.\label{digital}
\end{align}
Problem (\ref{digital}) has a well-known singular-value decomposition solution \cite{sixmethods, ICD}: 
\begin{align}
    \mathbf{F}[k] = \left(\left(e^{j\mathbf{\Phi}}\right)^*e^{j\mathbf{\Phi}}\right)^{-\frac{1}{2}}\frac{\mathbf{V}_{\text{eff}}[k]}{||\mathbf{V}_{\text{eff}}[k]||_F}\label{waterfilling_digital2},
\end{align}
where $\mathbf{V}_{\text{eff}}[k]$ is the truncated right singular vector matrix with columns corresponding to the largest $N_s$ nonzero singular values of $\mathbf{H}_{\text{eff}}[k]$.

\textit{2) Analog Beamformer Optimization Problem:} Now we focus on designing the analog beamformer only by fixing the digital beamformers $\mathbf{F}[k]$, and the problem in (\ref{prob_psi}) becomes:
{\setlength{\abovedisplayskip}{0pt}
 \setlength{\belowdisplayskip}{4pt}
\begin{align}
    \max_{\mathbf{\Phi}} \quad & \frac{1}{K}\sum_{k=1}^K \log_2[\det(\mathbf{I}_{N_r} + \frac{P_t}{\sigma_n^2}(\mathbf{H}[k]e^{j\mathbf{\Phi}}\mathbf{F}[k]\nonumber\\
    &\times \mathbf{F^*}[k]\left(e^{j\mathbf{\Phi}}\right)^*\mathbf{H^*}[k]))].\label{analog}
\end{align}
}The power constraint can be satisfied after solving (\ref{analog}) by renormalizing the digital beamformers as in (\ref{digital_reconstruct}).

\subsection{Message Generation and Node Update in AN-GNN}

To optimize the analog beamformer's phase $\mathbf{\Phi}$ in (\ref{analog}), we modify the NU-GNN structure introduced in Section \ref{sec:NUGNN} by using the closed form solution in (\ref{waterfilling_digital2}) to update the digital beamformer representations. This results in the AN-GNN structure in Fig. \ref{fig:AN_stru}. We design the learning and updating process of the AN-GNN as follows.

Before updating the messages at each layer $l$, we use the previous $\mathbf{x}^{(l-1)}$ to reconstruct the analog phase matrix $\mathbf{\Phi}^{(l-1)}$ as in (\ref{analog_reconstruct}), then use that together with each channel matrix $\mathbf{H}[k]$ to calculate the digital beamformer $\mathbf{F}[k]^{(l-1)}$ as in (\ref{waterfilling_digital2}). These digital beamformers are reshaped into digital beamformer representations $\mathbf{c}_k^{(l-1)}$ and used as input into the $l$-th GNN updating layer, as shown in orange in Fig. \ref{fig:AN_stru}. The feature at subcarrier node $k$ can be expressed as
{\setlength{\abovedisplayskip}{2pt}
 \setlength{\belowdisplayskip}{2pt}
\begin{align}
\mathbf{c}_k^{(l-1)}=\left[ \text{vec} \left( \text{Re} \left\{ \mathbf{F}[k] \right\} \right)^T, \text{vec} \left( \text{Im} \left\{ \mathbf{F}[k] \right\} \right)^T \right]^T.\label{node_feature}
\end{align}
}

In the $l$-th updating layer, the corresponding $\mathbf{c}_k$ is combined with the edge feature $\mathbf{h}_k$ to generate the message $\mathbf{m}^{(l)}_{k}$:
\begin{align}
    \mathbf{m}_{k}^{(l)}=f_1(\mathbf{h}_k,\mathbf{c}_k^{(l-1)})\label{AN:md},
\end{align}
where $\textit{f}_1(\cdot)$ is an MLP as shown in green in Fig. \ref{fig:AN_stru}. We update the analog representation as
{\setlength{\abovedisplayskip}{3pt}
 \setlength{\belowdisplayskip}{3pt}
\begin{align}
&\mathbf{x}^{(l)}=f_2(\mathbf{x}^{(l-1)}, g({\mathbf{m}_{k}^{(l)}})_{k\in\mathcal{N}(\mathbf{x})})\label{AN:x},
\end{align}}where $\textit{f}_2(\cdot)$ is an MLP marked in blue as shown in Fig. \ref{fig:AN_stru}. $g(\cdot)$ is the attention aggregation as discussed in the next subsection.

\subsection{Aggregations Via an Attention Mechanism}

\begin{figure}[!t]
\hspace{-0.06\linewidth}
\centering
\includegraphics[width=2.6in]{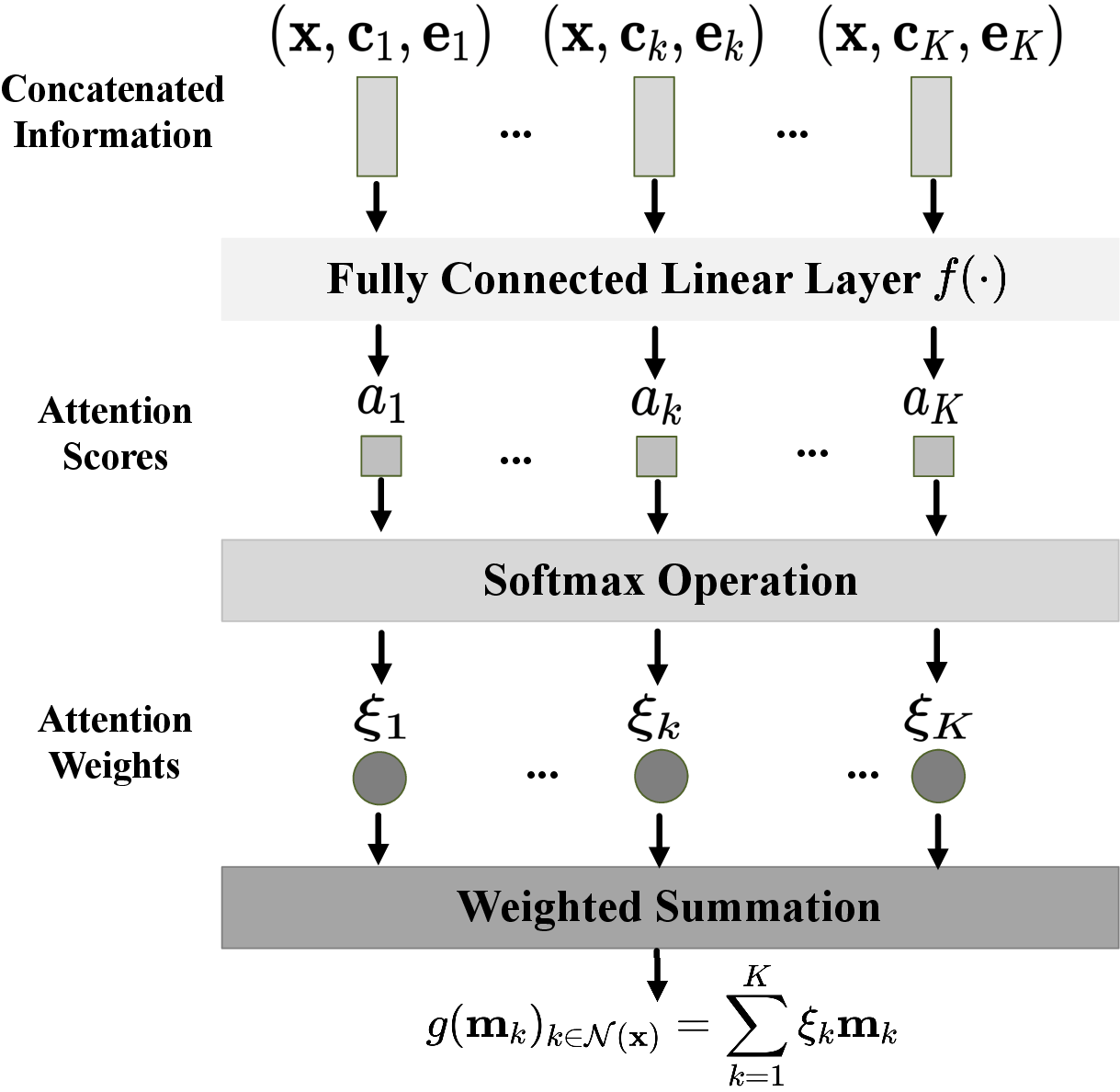}
\caption{The attention aggregation approach: First a fully connected linear layer is used to calculate the attention score $a_k$ for each subcarrier, then these scores are converted into attention weights $\xi_k$ by applying a softmax operation. A weighted sum by the learned $\xi_k$ values emphasizes the relative importance of different subcarriers.}
\vspace{-7pt}
\label{fig_sa}
\end{figure}

An important change compared to previous GNN structures is that in the updating processes in (\ref{AN:x}), $g(\cdot)$ is no longer the element-wise mean aggregation function. Here, we draw inspiration from the attention mechanism to better differentiate the impact of individual subcarriers on the analog beamformer. Specifically, we use a fully-connected linear layer followed by a LeakyReLU activation to calculate the attention score for each subcarrier. Then, we apply a softmax operation to convert these scores into attention weights. By performing a weighted sum, the relative importance of information from different neighboring nodes representing different subcarriers is adaptively adjusted for aggregation. The attention weights for each subcarrier can be calculated as:
{\setlength{\abovedisplayskip}{4pt}
 \setlength{\belowdisplayskip}{4pt}
\begin{align}
    \xi_k = \frac{\text{exp}\left(\text{LeakyReLU}\left(f(\mathbf{x}, \mathbf{c}_k,\mathbf{h}_k)\right)\right)}{\sum_{k=1}^K\text{exp}\left(\text{LeakyReLU}\left(f(\mathbf{x}, \mathbf{c}_k,\mathbf{h}_k)\right)\right)},\label{attention1}
\end{align}}where $f(\cdot)$ is a linear layer. The aggregated information is
{\setlength{\abovedisplayskip}{2pt}
 \setlength{\belowdisplayskip}{4pt}
\begin{align}
g(\mathbf{m}_k)_{k\in\mathcal{N}(\mathbf{x})} = \sum_{k=1}^K \xi_k\mathbf{m}_k.\label{attention2}
\end{align}
}The overall process is shown in Fig. \ref{fig_sa}. This attention-based aggregation is permutation invariant. Indeed, the attention scores in Eq. (\ref{attention1}) are computed via a permutation equivariant activation, followed by a softmax function that preserves this property. Consequently, both $\xi_k$ and $\mathbf{m}_k$ in (\ref{attention2}) are jointly permutation invariant under subcarrier reordering, leading to a permutation invariant outcome.

\subsection{Beamformer Reconstruction}
We employ (\ref{analog_reconstruct}) to reconstruct the analog beamforming phase matrix $\mathbf{\Phi}^{(L)}$ from the analog node representation $\mathbf{x}^{(L)}$. We then use $\mathbf{\Phi}^{(L)}$ to update the digital beamforming matrices as in (\ref{waterfilling_digital2}) and (\ref{digital_reconstruct}). The resulting $\mathbf{\Phi}^{(L)}$ and $\mathbf{F}_{\text{normalized}}[k]$ are used to compute the sum rate in (28) for both training and inference. The overall algorithm is summarized in Alg. \ref{AN_algo}.

\setlength{\textfloatsep}{4pt}
\renewcommand{\algorithmicrequire}{\textbf{Input:}}
\begin{algorithm}
\caption{AN-GNN with Attention Aggregation}
\label{AN_algo}
\begin{algorithmic}[1]
\Require $\mathbf{H}[k]$, where $k=1,\ldots,K$ 
\For{each epoch}
    \For{each sample $\in \mathcal{B}$}
        \State Generate the edge feature $\mathbf{h}_k$ as shown in (\ref{edge}).
        \State Initialize $\mathbf{x}^{(0)}$ and $\mathbf{m}^{(0)}$.
        \For{$l=1$ to $L$}
            \State Reconstruct $\mathbf{\Phi}^{(l-1)}$ by using $\mathbf{x}^{(l-1)}$.
            \State Calculate $\mathbf{F}^{(l-1)}[k]$ by (\ref{waterfilling_digital2}).
            \State Generate $\mathbf{c}^{(l-1)}_k$ as shown in (\ref{node_feature}).
            \State Perform forward propagation as in (\ref{AN:md})--(\ref{attention2}).
        \EndFor
        \State Reconstruct $\mathbf{\Phi}^{(L)}$ by using $\mathbf{x}^{(L)}$.
        \State Calculate $\mathbf{F}_{\text{normalized}}[k]$ by (\ref{waterfilling_digital2}) and (\ref{digital_reconstruct}).
        \State Calculate the loss function according to (\ref{loss}).
    \EndFor
    \State Update the GNN's parameters according to (\ref{minibatch}) for the next epoch.
\EndFor
\end{algorithmic}
\end{algorithm}

\section{Extension to the Multi-User Setting}\label{section7}
In this section, we directly extend the two pure machine learning approaches proposed in Section \ref{section5} (NU-GNN and EU-GNN) to the multi-user case. This extension is achieved by reformulating the objective function to account for multi-user transmission and expanding the graph model to include multiple users while preserving the same embedded information and GNN updating rules. The AN-GNN cannot be directly extended to the multi-user case, however, because the digital beamformers have no closed-form solution for the multi-user setting as in the single-user scenario.

\subsection{Problem Reformulation}
We consider the same wideband MIMO-OFDM hybrid beamforming system model as in Section \ref{section2}, and extend it to a multi-user scenario directly. In this case, the BS serves $M$ UEs and transmits $N_d$ data streams to each UE, resulting in a total of $N_s=MN_d$ data streams. The transmitted signal vector on the $k$-th subcarrier is
{\setlength{\abovedisplayskip}{4pt}
 \setlength{\belowdisplayskip}{4pt}
\begin{align}
    \mathbf{x}[k] =\sqrt{P_t}e^{j\mathbf{\Phi}}\sum_{m=1}^M\mathbf{F}_m[k]\mathbf{s}_m[k], \label{trans_sig}
\end{align}
}where $e^{j\mathbf{\Phi}} \in \mathbb{C}^{N_t \times N_{\text{RF}}}$ is the analog beamformer with the constant modulus constraint as in Eq. (\ref{modulo_constraint2}). $\mathbf{F}_m[k] \in \mathbb{C}^{N_{\text{RF}} \times N_d}$ is the digital beamformer for UE $m$ and subcarrier $k$.

The transmit power constraint per subcarrier remains the same as in (\ref{power_constraint}), where $P_t$ denotes the total transmit power allocated to each subcarrier across all UEs. Therefore, the digital beamformers implicitly handle the power allocation among different users.

The received signal for the $m$-th user on the $k$-th subcarrier is
\begin{align}
\mathbf{y}_m[k] &= \underbrace{\sqrt{P_t}\mathbf{H}_m[k]e^{j\mathbf{\Phi}}\mathbf{F}_m[k]\mathbf{s}_m[k]}_{\text{desired signal}}\nonumber
    \\[1pt]
    &+ 
    \underbrace{
    \sqrt{P_t} \sum_{\substack{n=1 \\ n \neq m}}^M \mathbf{H}_m[k]e^{j\mathbf{\Phi}}\mathbf{F}_n[k]\mathbf{s}_n[k]
    }_{\text{inter-user interference}} +{\underbrace{\mathbf{n}_m[k]}_{\text{noise}}}, \label{multi_user_received_signal}
\end{align}
where $\mathbf{n}_m[k]\sim\mathcal{CN}(0,\sigma^2\mathbf{I}_{N_r})$ is the additive white Gaussian noise for each UE $m$.

Based on the received signal model in (\ref{multi_user_received_signal}), the achievable spectral efficiency can be expressed as
\begin{align}
    R_{\text{MU}} = \frac{1}{K}\sum_{k=1}^K \sum_{m=1}^M&\text{log}_2 \biggl[\det\biggl(\mathbf{I}_{N_r} + P_t\mathbf{H}_m[k]e^{j\mathbf{\Phi}}\mathbf{F}_m[k] \nonumber\\ &\times \mathbf{F}_m^*[k](e^{j\mathbf{\Phi}})^*\mathbf{H}_m^*[k]\times \mathbf{\Gamma}_\text{inter}^{-1}\biggr)\biggr],\label{SE_multi_user}
\end{align}
where $\mathbf{\Gamma}_\text{inter}$ is the interference-plus-noise covariance matrix:
\begin{align}
    \mathbf{\Gamma}_\text{inter} = P_t\sum_{\substack{n=1 \\ n \neq m}}^M \mathbf{H}_m[k]e^{j\mathbf{\Phi}}\mathbf{F}_n[k]\mathbf{F}_n^*[k](e^{j\mathbf{\Phi}})^*\mathbf{H}^*_m[k] + \sigma^2 \mathbf{I}_{N_r}.
\end{align}

Subsequently, the reformulation in (\ref{prob_psi}) still applies by replacing the objective function with (\ref{SE_multi_user}). This multi-user objective will be used to construct the training loss for the
proposed GNNs in (\ref{loss}).

\subsection{Extended Graph Model and Updating Rules}
\begin{figure}[!t]
\centering
\includegraphics[width=2.8in]{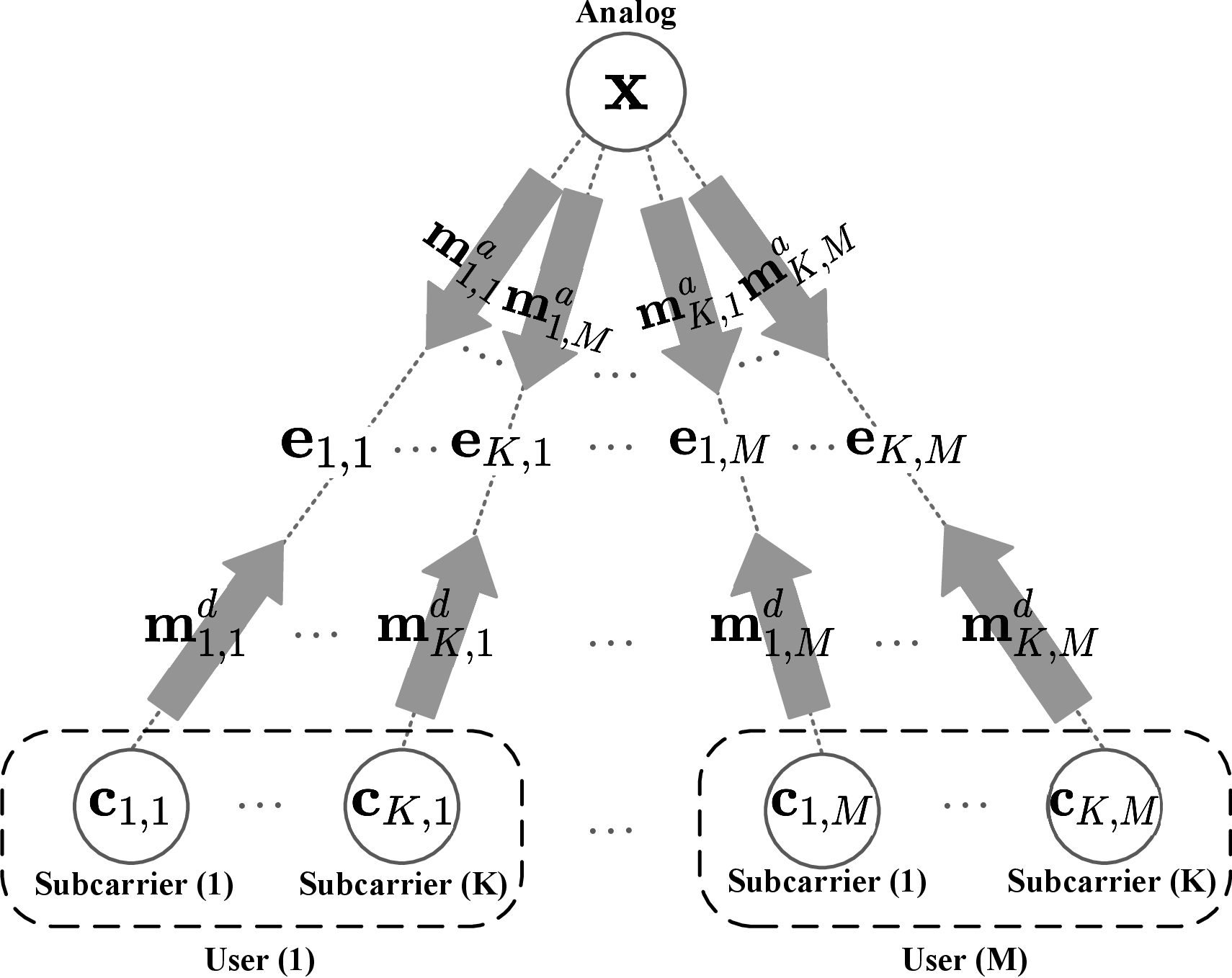}
\caption{Direct extension to multi-user case using the same bipartite graph model with two types of nodes: analog and subcarrier nodes. Information embedded at the analog node, each subcarrier node, and edge is represented as vector $\mathbf{x}$, $\mathbf{c}_{k,m}$, and $\mathbf{e}_{k,m}$, respectively. Each dashed box corresponds to a user-specific subcarrier set, which groups all subcarrier nodes and their associated edges corresponding to a given user $m$.}
\vspace{-2pt}
\label{fig:graph_model_MU}
\end{figure}

Next, we extend the graph model and
updating rules of the proposed NU-GNN and EU-GNN to the multi-user scenario.

\textit{1) Graph Model:} We extend the graph model in Fig.~\ref{fig:graph_model} to the multi-user case by introducing additional subcarrier nodes for new users, as illustrated in Fig.~\ref{fig:graph_model_MU}. Specifically, the subcarrier nodes corresponding to each user are grouped into a user-specific subcarrier set. This extension retains the permutation equivariance property of the proposed GNN, as stated in Proposition 3.

\textit{Proposition 3:} The output of each GNN updating layer remains permutation equivariant with respect to both the subcarrier and user orders under the proposed multi-user graph extension.

\textit{Proof:} See Appendix~C. \hfill $\blacksquare$

Proposition 3 implies generalization across varying subcarrier and user orders without retraining.

\textit{2) GNN Updating Rules:} Since the underlying graph model is similar to the single-user case (except with a different loss function that dictates the interaction among nodes), the updating rules of the proposed NU-GNN and EU-GNN are directly applicable to the multi-user case. Accordingly, the message generation and representation update operations follow the same rules as in the single-user setting in Eqs. (\ref{NU_ma})--(\ref{EU:e}). The only difference lies in the node indexing, where the subcarrier index $k$ is extended to $(k,m)$ to account for the $m$-th user. Accordingly, the neighboring set is extended to $(k,m) \in \mathcal{N}(\mathbf{x})$ during aggregation, through which the inter-user interference is naturally captured from neighboring nodes and directly accounted for in the loss function.

\textit{3) Beamformer Reconstruction:} To satisfy the transmit power constraint in (\ref{prob_psi}), we first reconstruct the digital beamformer for each user $m$ on each subcarrier $k$. Specifically, $\mathbf{F}_{k,m}=\text{reshape}\!\left(\mathbf{c}_{k,m}^{(L)}\left[1:N_{\text{RF}}\times N_d\right],\left(N_{\text{RF}},N_d \right) \right) + j\ \text{reshape}\!\left(\mathbf{c}_{k,m}^{(L)}\left[N_{\text{RF}}\times N_s:2N_{\text{RF}}\times N_d\right],\left(N_{\text{RF}},N_d\right)\right)$. 
The reconstructed digital beamformers are then concatenated to form $\mathbf{F}_k=\begin{bmatrix}
\mathbf{F}_{k,1} &
\cdots&
\mathbf{F}_{k,M} 
\end{bmatrix}$. The resulting digital beamformer is subsequently normalized according to (\ref{digital_reconstruct}), and the reconstruction of the analog beamformer follows the same procedure as in (\ref{analog_reconstruct}).

\section{Computational Complexity Analysis}\label{section8}
In this section, we provide an analysis of the forward computation complexity of our three GNN models during inference, and compare them with that of a traditional optimization method, the AMO algorithm in \cite{AMO}. Since the training of the GNNs can be done offline before the models are deployed, we do not analyze training complexity. Instead, we focus on inference complexity since it captures the runtime computational requirements and is relevant for applying the trained GNNs in a practical communication system.

\subsection{Proposed GNN Structures}
Our NU-GNN and EU-GNN structures both consist of 4 different MLPs of the same depth but different layer sizes. The computational complexity of the $i$-th MLP depends on the number of hidden layers $d$, and the size of each layer including input size $n_i$, each hidden layer size $2n_i$, and output size $m_i$. The forward computational complexity can be calculated as $O\left(2n_i^2+ 4dn_i^2+2n_im_i\right)$. Typically, $n_i \gg m_i$ in each MLP, thus the complexity order per MLP can be simplified as $O\left(dn_i^2\right)$. During the forward propagation process, MLP $\textit{f}_2^{\,a}(\cdot)$ is computed once, while the other three are computed $K$ times. Therefore, we can express the overall complexity for one GNN updating layer as $O\left(dKn^2\right)$, where $n=\max\{n_i, i\in[1,4]\}$.


AN-GNN shares a similar structure with NU-GNN but with an aggregation that includes a fully connected linear layer, contributing a computational complexity of $O\left(2N_tN_r+N_tN_{\text{RF}}+2N_{\text{RF}}N_s\right)$. Unlike NU-GNN, the subcarrier node features in the AN-GNN are computed not through MLPs but via matrix multiplication and performing the SVD. The overall forward computation complexity is $O\left(dKn^2\right)+O\left(2N_tN_r+N_tN_{\text{RF}}+2N_{\text{RF}}N_s\right)+O(N_tN_{\text{RF}}+KN_rN_tN_{\text{RF}}+KN_r^2N_{\text{RF}}^{t})$, where the first term remains the dominant one.

Given the above complexity order for each GNN updating layer, the total complexity for all $L$ layers of a GNN is $O\left(dLKn^2\right)$. For the beamformer reconstruction operations in (\ref{analog_reconstruct}) and (\ref{digital_reconstruct}), the element-wise exponential operation on a matrix requires $O(N_tN_{\text{RF}})$ complexity, while the Frobenius norm and matrix multiplication requires $O(KN_tN_{\text{RF}}N_s)$. Among all these complexities, the dominant term is still the calculation of $L$ updating layers. As a result, the overall computational complexity for processing one sample through the GNN can be expressed as $O\left(dLKn^2\right)$, where $n=2N_tN_r+N_tN_{\text{RF}}$ for the NU-GNN and AN-GNN, and $n=2N_tN_r++N_tN_{\text{RF}}+2N_{\text{RF}}N_s$ for the EU-GNN. Since $N_t \gg N_{\text{RF}} \geq N_s$ and $N_r > N_s$, for all cases, the complexity of processing one GNN sample can be simplified as 
{\setlength{\abovedisplayskip}{4pt}
 \setlength{\belowdisplayskip}{4pt}
\begin{align}
    O( d L K N_t^2 N_r^2 )\label{GNN_O},
\end{align}
}where $d$ and $L$ are GNN settings, and $K$, $N_t$ and $N_r$ are wireless network settings.

\subsection{AMO Algorithm \cite{AMO}}
For the AMO algorithm in \cite{AMO}, the computational steps involve manifold optimization (MO) to update the value of $\mathbf{W}$. According to Algorithm 1 in \cite{AMO}, the key computational costs are from the Armijo backtracking line search, retraction, cost function, and the computation of the Riemannian gradient. Assume that an average of $J$ iterations is required to find an appropriate step size using the Armijo method. The complexity of computing the Riemannian gradient at each update step along with the cost function computation is $O\left(3JK[N_t^3N_s^2N_{\text{RF}}]+2JK[N_t^3(N_{\text{RF}})^2N_s]+5JN_t^2(N_{\text{RF}})^2\right)$. Since $N_s \leq N_{\text{RF}}$, the second term is dominant.

In each Riemannian update step, assume an average of $I$ iterations for $\mathbf{W}$ to converge to the current optimal solution given $\mathbf{F}[k]$. The complexity of retraction, which normalizes the amplitude of elements in $\mathbf{W}$ to 1, is $O(IN_tN_{\text{RF}})$, which is negligible compared to the line search. Furthermore, computing the $K$ beamformers $\mathbf{F}[k]$ requires matrix inversion and multiplication. Combined with the power constraint, the resulting complexity is $O\left(K(N_{\text{RF}})^3+2KN_tN_{\text{RF}}N_s\right)$, which remains minor since $N_{\text{RF}} \ll N_t$.

Therefore, the overall complexity of the AMO algorithm, which requires an average of $M$ iterations of Riemannian updates to converge, can be approximated as:
{\setlength{\abovedisplayskip}{4pt}
 \setlength{\belowdisplayskip}{4pt}
\begin{align}
O\left(MIJKN_t^3(N_{\text{RF}})^2N_s\right)
,\label{AMO_O}
\end{align}
}where $J$ is the number of iterations required for each line search to find the step size, $I$ for the current optimal value of $\mathbf{W}$, and $M$ for the convergence of the algorithm.

\subsection{Comparison Between GNNs and AMO}
Since $N_t \gg N_{\text{RF}} \geq N_s$ and $N_r \geq N_s$, when comparing the complexities of the two algorithms, the dominant term is the highest degree term related to $N_t$. From (\ref{GNN_O}), we observe that the complexity of a GNN pass is of second order in $N_t$, whereas for AMO in (\ref{AMO_O}), it reaches cubic. This indicates that as we increase the number of antennas, especially in massive MIMO systems, the GNN model offers a significant advantage in terms of computation cost savings. In addition, as we increase the number of subcarriers $K$ in an OFDM system, the complexity of both algorithms grows linearly. However, the multiplier factor associated with $KN_t^3$ in (\ref{AMO_O}) is much larger than the one associated with $KN_t^2$ in (\ref{GNN_O}). This implies that the complexity of the AMO algorithm will increase much faster than that of our proposed GNN with respect to the transmit antenna array size and number of subcarriers.

\section{Numerical Simulations}\label{section9}

\subsection{Simulation System Settings}
Here we describe the settings for the simulation system and for training the GNN structures. 

\textit{1) Communication System Settings:} We use a carrier frequency of $f_c=142$GHz and a bandwidth of $B=20$GHz, with $K=4$ subcarriers selected for the offline training process. The number of subcarriers will be varied during testing and evaluation. The BS employs an $N_t = 64$ UPA antenna system, equipped with $N_s=N_{\text{RF}}=4$ RF chains, while the UE uses an $N_r = 8$ UPA antenna system. {In all simulations, the per-subcarrier transmit power is set to be equal to $P_t$, except for one result in Fig. \ref{fig:inf_NU} (noted with $\leq P_t$).}

The channel generation follows the measurement-based mmWave double directional model in \cite{channel}, which includes both large and small scale fading effects. The channel parameters are defined with $N_{\text{cl}}=2$ clusters and $N_{\text{ray}}=3$ rays per cluster. The complex path gain $\alpha_{il}=a_{il}e^{j\psi_{il}}$ is a product of the real large-scale fading amplitude $a_{il}$ and the channel phase component $\psi_{il} \in [0,2\pi)$. Here $a_{il}$ follows the probabilistic-based pathloss model described in \cite{channel, probLoS}, with an averaged transmit power $P_t=36$dBm and a noise power spectral density of $-174$dBm/Hz. The UE distribution is followed by a uniform location distribution within a circular area, with a TX-RX distance ranging in [10,100]m. The propagation path delay follows $\tau_{il} \sim \mathcal{U}(0,100\text{ns}) $\cite{channel}. Both the azimuth and elevation angles are following a wrapped Gaussian distribution \cite{channel, NYUSIM}. The antenna elements are spaced at $d=\frac{\lambda_c}{2}$.

\textit{2) Machine Learning Settings:} During offline training, we initialized the GNN as $\mathbf{x}\sim\mathcal{U}[0,2\pi)$, $\mathbf{c}_k\sim\mathcal{N}(0,1)$, and $\mathbf{e}_k$ as in (\ref{edge}). All MLPs had two hidden layers, each with the number of neurons twice the MLP input size, followed by ReLU activations. For EU-GNN, we applied dropout (30$\%$) to hidden layers during message generation to improve convergence and generalization. We used the Adam optimizer with a $5\times 10^{-4}$ learning rate, halved every 200 epochs for NU-GNN and EU-GNN. This value was empirically selected as the best within the range of $[10^{-5}, 10^{-3}]$. AN-GNN used a warm restart scheduler \cite{warmrestart}, cyclically sweeping the learning rate in $[5\times 10^{-5}, 5\times 10^{-4}]$ to enhance performance. Training used mini-batches of 100 samples, with 100 batches per epoch. All three GNN models have $L=2$ layers, as the bipartite graph can be fully traversed in two hops. This choice of $L$ also helps mitigate the oversmoothing effect \cite{oversmoothing}, which degrades the ability of different nodes to converge to distinct representations. We observed such effects in early experiments when using $L=4$.

\textit{3) Baseline Schemes:} We compare our proposed GNNs with the following baseline schemes:
\begin{itemize}[leftmargin=10pt]
    \item \textbf{FD:} Fully digital beamforming method, which applies SVD to the channel matrix of each subcarrier to obtain the optimal fully digital beamformer.
    \item \textbf{AMO \cite{AMO}:} Iteratively updating the analog beamformer via manifold optimization and alternating with digital beamforming update.
    \item \textbf{ICD \cite{ICD}:} Iteratively updating the analog beamformer via coordinate descent and alternating with digital beamformers.
    \item \textbf{AV-all \cite{sixmethods}:} Averaging the array response vectors across all subcarriers to construct the analog beamformer.
    \item \textbf{MCM \cite{sixmethods}:} Using dominant eigenvectors of the mean channel matrix across all subcarriers for analog beamforming.
    \item \textbf{FNN:} A fully-connected neural network (FNN) is constructed to learn both the analog and digital beamformers.
    \item \textbf{AN-FNN:} A fully-connected neural network (FNN) is constructed to learn only the analog beamformer.
    \item \textbf{LCMLP-GNN \cite{Yang2024FDD}:} A GNN method using a linear combination of the learned representations without message-passing as updating layers. 
\end{itemize}

\subsection{ML Performance and Complexity Analysis}
\begin{figure}[!t]
    \centering
    \subfloat[]{%
        \includegraphics[width=0.43\linewidth]{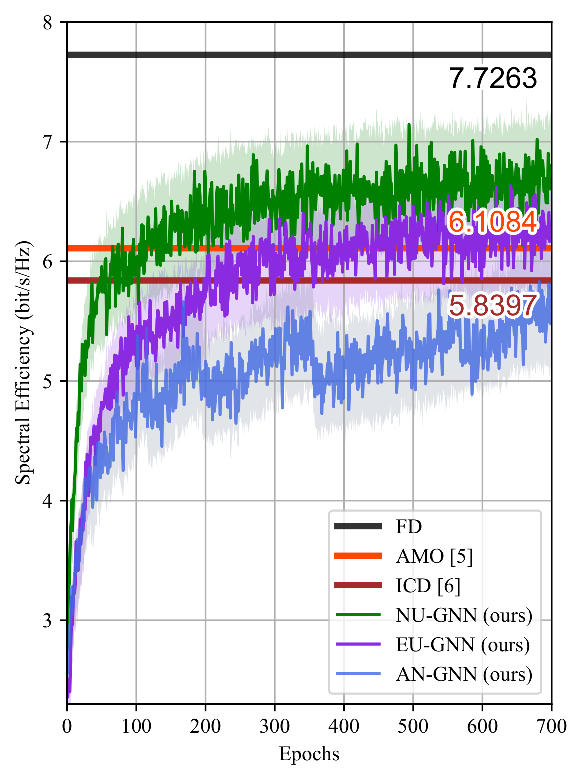}%
        \label{fig:Conv_all}%
    }
    \subfloat[]{%
        \includegraphics[width=0.43\linewidth]{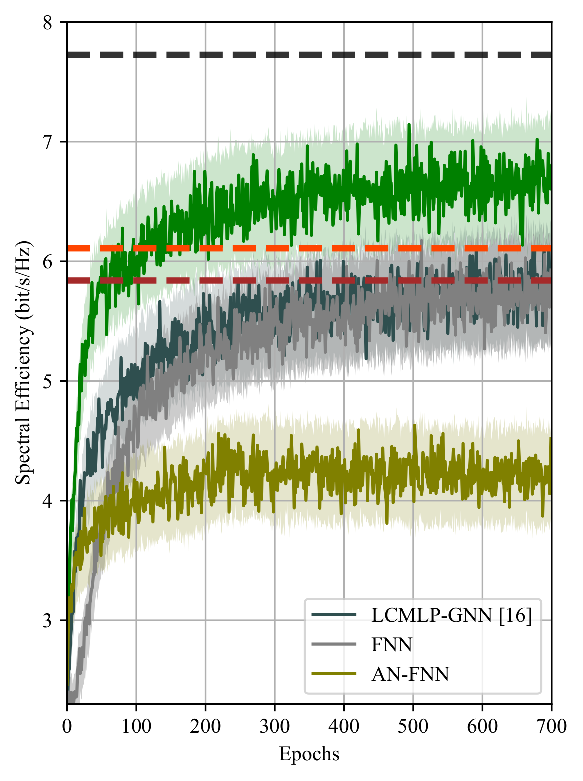}%
        \label{fig:Conv_NU}%
    }
    \caption{Convergence of the proposed GNNs training for $P_t$ = 36dBm, $N_t=64$, $K=4$, $f_c=142$GHz, and $B=20$GHz. Training curves are shown as shaded regions, while solid lines indicate validation curves. (a) Comparison among the three proposed GNN structures, with traditional methods and FD. (b) Comparison of NU-GNN with three other ML benchmark methods.}
    \vspace{-1pt}
    \label{fig:training}
\end{figure}

\textit{1) Data Rate Convergence:} Fig. \ref{fig:training} shows the convergence of the proposed GNN structures. The NU-GNN outperforms the AMO benchmark and also outperforms all the other ML benchmark methods. Additionally, the EU-GNN converges to a spectral efficiency approaching that of the AMO, and the AN-GNN achieves performance similar to that of the ICD.

\begin{table*}[!t]
\renewcommand{\arraystretch}{1.3}
\caption{Ablation study of proposed GNN variants. See the text for the detailed description of different variants.}
\label{ablation}
\centering
{
\begin{tabular}{|
>{\centering\arraybackslash}p{2.5cm}||
>{\centering\arraybackslash}p{1.5cm}|
>{\centering\arraybackslash}p{1.5cm}|
>{\centering\arraybackslash}p{1cm}|
>{\centering\arraybackslash}p{1.2cm}|
>{\centering\arraybackslash}p{1.0cm}|
>{\centering\arraybackslash}p{1.0cm}|
>{\centering\arraybackslash}p{1.0cm}|
>{\centering\arraybackslash}p{1.3cm}|
>{\centering\arraybackslash}p{1.3cm}|
}
\hline
\multirow{3}{*}{\textbf{Methods}} 
& \multirow{3}{*}{\makecell{\textbf{Converged} \\ \textbf{Values} \\  \textbf{(bps/Hz)}}} 
& \multicolumn{3}{c|}{\textbf{Training Time}} 
& \multicolumn{2}{c|}{\textbf{Inference Time}} 
& \multicolumn{2}{c|}{\textbf{Storage}} 
& \multirow{3}{*}{\textbf{\# Params}} \\
\cline{3-9}
& & \multirow{2}{*}{\makecell{\textbf{Mean/Epoch} \\ \textbf{(sec)}}}
  & \multirow{2}{*}{\makecell{\textbf{Std} \\ \textbf{(sec)}}}
  & \multirow{2}{*}{\makecell{\textbf{Total} \\ \textbf{(min)}}}
  & \multirow{2}{*}{\makecell{\textbf{Mean} \\ \textbf{(sec)}}}
  & \multirow{2}{*}{\makecell{\textbf{Std} \\ \textbf{(sec)}}}
  & \multirow{2}{*}{\makecell{\textbf{Mean} \\ \textbf{(Mb)}}}
  & \multirow{2}{*}{\makecell{\textbf{Std} \\ \textbf{(Mb)}}}
  & \\ & & & & & & & & & \\
\hline
\textbf{NU-GNN} & $\mathbf{6.7533}$ & 205.19 & 5.21 & 2386.43 & $\mathbf{0.0866}$ & 0.0006 & 567.97 & $1.5 \times 10^{-5}$  & 60.202M\\
\hline
NU-GNN (ExtMsg)  & 6.3426 & 230.09 & 11.87 & 2684.38 & 0.0974 & 0.0009 & 634.45 & $1.5 \times 10^{-5}$ & 74.054M\\
\hline
\textbf{EU-GNN} & 6.2409 & 234.03 & $\mathbf{4.11}$ & 2745.45 & 0.0907 & $\mathbf{0.0005}$ & 773.64 & $1.5 \times 10^{-5}$ & 133.369M \\
\hline
EU-GNN (add $\mathbf{h}$) & 6.3248 & 244.39 & 6.37 & 2882.98 & 0.0945 & 0.0008 & 1014.53 & $1.5 \times 10^{-5}$ & 196.427M \\
\hline
EU-GNN (ExtMsg) & 6.4022 & 269.96 & 8.09 & 3179.65 & 0.0986 & 0.0006 & 890.55 & $1.5 \times 10^{-5}$ & 163.803M\\
\hline
\textbf{AN-GNN} & 5.7004 & 177.09 & 6.7894 & 2078.56 & 0.1174 & 0.0008 & 356.28 & $1.5 \times 10^{-5}$ & 24.411M\\
\hline
AN-GNN (No Attn) & 5.6653 & $\mathbf{161.62}$ & 4.76 & $\mathbf{1905.48}$ & 0.0971 & 0.0008 & $\mathbf{356.27}$ & $1.5 \times 10^{-5}$ & \textbf{24.408M}\\
\hline
AMO \cite{AMO} & 6.1084 & -- & -- & -- & 2.8991 & 1.6632 & 4677.82 & 1846.4 & - \\
\hline
ICD \cite{ICD} & 5.8397 & -- & -- & -- & 0.3681 & 0.2798 & 53.7 & 3.8 & - \\
\hline
\end{tabular}
}
\vspace{-7pt}
\end{table*}

\textit{2) Ablation Study:} In our ablation study, we analyze the effect of the message and representation update methods, the choice of information embedding across all our proposed GNN structures, and the impact of attention specifically in AN-GNN. The comparison results across these GNN variants are summarized in Table \ref{ablation}.

First, we compare our designed efficient structure with a more complex message-passing structure. NU-GNN (ExtMsg) incorporates extra messages by additionally concatenating $\mathbf{m}^{d(l-1)}_k$ in the inputs of generating $\mathbf{m}^{a(l)}_k$ in Eq. (\ref{NU_ma}), and similarly includes $\mathbf{m}^{a(l-1)}_k$ in Eq. (\ref{NU_md}). Although this seems to provide more information during representation updates, the information carried by the additional messages is already contained in our simpler message-passing structure, and hence is double-counted, introducing unnecessary redundancy and complexity without increasing the convergence values.

For EU-GNN, we observe that EU-GNN (ExtMsg) improves the final converged value slightly, possibly because in EU-GNN, the CSI information is gradually overwritten during updates, causing the process to rely more on the graph structure than the input features. This motivated us to test EU-GNN ($\mathbf{h}_k$) by including the CSI inputs $\mathbf{h}_k$ as additional edge features in Eq.~(\ref{edge}), which similarly improves convergence values slightly. However, as the overall performance remains comparable while training time increases, we ultimately chose the simplest EU-GNN variant to balance performance and computational efficiency.

Finally, we explored the impact of incorporating attention-based aggregation. The AN-GNN with attention achieves a modest improvement in converged value compared to the other at a small increase in training time. Since the training of the AN-GNN with attention is still smaller than NU-GNN and EU-GNN, we selected this attention-based AN-GNN since it offers superior generalization ability, as shown in Fig. \ref{fig_generalization}.



\textit{3) Offline Training Time:} {Table \ref{ablation} provides the total training time, per-epoch mean training time, standard deviation, and trainable parameters for the three proposed GNN models. Taking the largest model EU-GNN as the baseline, AN-GNN reduces parameters by 80\% and speeds up training by 25\%, while NU-GNN achieves a 59\% parameter reduction and a 13\% speed-up. AN-GNN's shorter mean training time offers a meaningful advantage if the training is done online, where only fine-tuning is needed for new incoming data. EU-GNN exhibits the lowest standard deviation, indicating more consistent updates and greater training stability. On the other hand, AN-GNN shows the highest standard deviation, primarily due to the SVD computations in the digital beamformers at each algorithm step, which introduce runtime fluctuations.



\textit{4) Online Inference Running Time:} The practicality of our GNNs is highlighted in the inference time comparison in Table \ref{ablation}. Across $10^3$ channel realizations, AMO is over 32 times or more than an order of magnitude slower than the proposed GNNs. This is because AMO needs to repeat the alternative optimization process for each subcarrier and channel realization, while the GNNs simply utilize the pre-trained models to perform feed-forward computation, producing results quickly by directly scaling up the number of subcarrier nodes. Furthermore, the GNNs exhibit computation time standard deviations three orders of magnitude lower than that of AMO, ensuring highly stable computation time per CSI update. Notably, although AN-GNN has the fewest trainable parameters, this advantage is compromised by the SVD and matrix inversion in computing the digital beamformers, which ultimately leads to a slower inference time.


\textit{5) Dynamic Memory Allocation:} Table \ref{ablation} further emphasizes the GNNs' computational efficiency in memory use. Since we use the pre-trained GNN models directly during the online inference phase, the amount of dynamic memory required for each channel realization remains constant, where EU-GNN requires the most dynamic memory allocation, and AN-GNN the least. In contrast, AMO's repeated optimization leads to high variability and an average memory usage nearly 8 times higher than that of the GNN models. This highlights GNNs' efficiency and stability in resource allocation, making them more suitable for practical deployment.

\subsection{Communication Performance}
During the online inference phase for simulation, we increased the number of subcarriers to $K=64$, and averaged all presented simulation results over $10^3$ channel realizations.

\begin{figure}[!t]
    \centering
    \subfloat[]{%
        \includegraphics[width=0.4\linewidth]{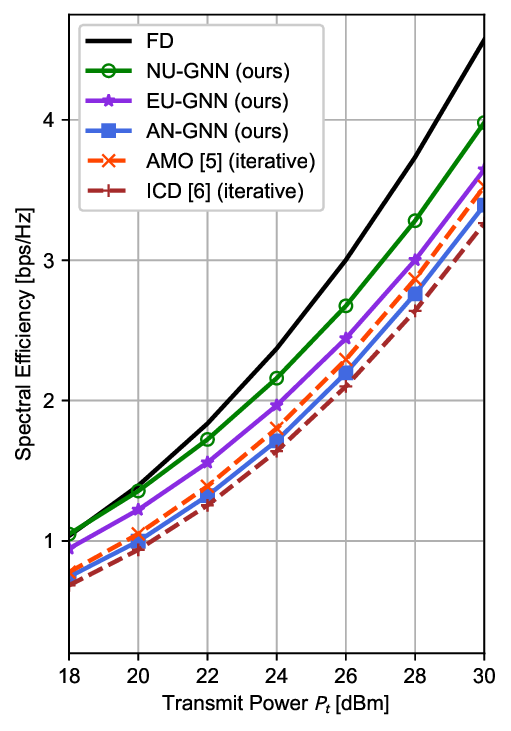}
    \label{fig:inf_ours}
    }
    \subfloat[]{%
        \includegraphics[width=0.4\linewidth]{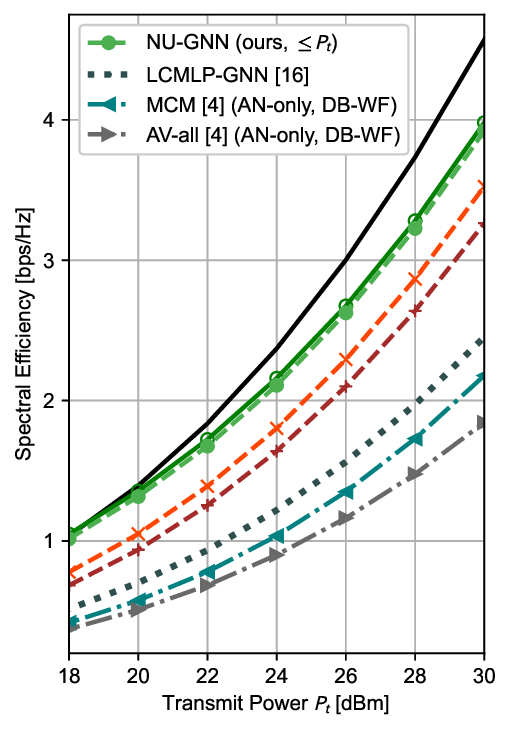}%
        \label{fig:inf_NU}%
    }
    \caption{Spectral efficiency achieved by different beamforming design algorithms with $K=64$ subcarriers, averaged over $10^3$ channel realizations. (a) Comparison among the three proposed GNN structures, with FD and traditional methods. (b) Comparison of NU-GNN with three other benchmarks.}
    \label{fig_run_n10_0}
    \vspace{-2pt}
\end{figure}

\textit{1) Spectral Efficiency vs $P_t$:} Fig. \ref{fig_run_n10_0} shows that our NU-GNN and EU-GNN outperform all traditional methods, including the two iterative optimization algorithms (AMO \cite{AMO} and ICD \cite{ICD}). Among the methods that all use singular-value decomposition closed form to calculate the digital beamformers (ICD \cite{ICD}, MCM \cite{sixmethods}, AV-all \cite{sixmethods}), our AN-GNN achieves the best performance. {In addition, as shown in Fig. \ref{fig:inf_NU}, the performance difference between equal and upper-bounded subcarrier power allocation is relatively small, indicating that the observed performance gain mainly comes from the proposed GNN-based hybrid beamforming design rather than the specific power normalization strategy.}

\begin{figure}[!t]
\centering
\includegraphics[width=0.7\linewidth]{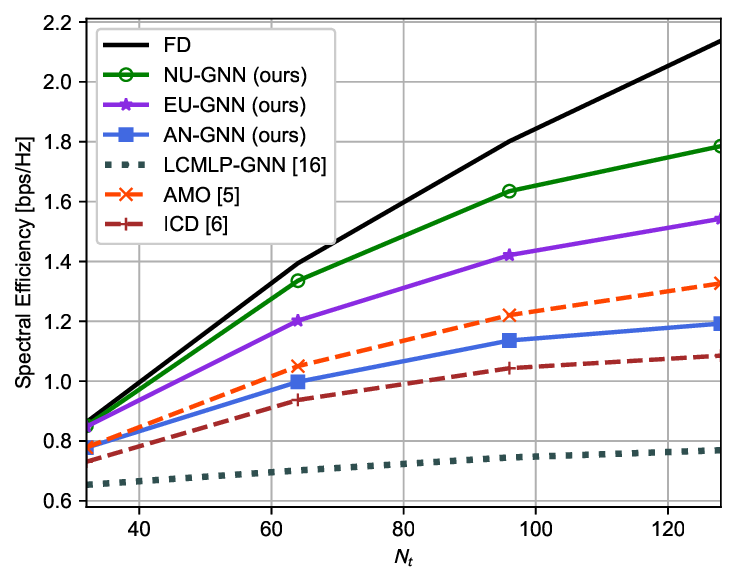}
\vspace{-5pt}
\caption{Spectral efficiency versus the number of transmitter antennas $N_t$, with $K=64$ subcarriers, $P_t=20$dBm, averaged over $10^3$ channel realizations.}
\vspace{-2pt}
\label{fig_multiantennas}
\end{figure}

\textit{2) Spectral Efficiency vs Antenna Array Size:} We evaluated our proposed GNN models across different antenna array sizes by training additional GNNs with different $N_t$. Shown in Fig. \ref{fig_multiantennas}, as $N_t$ increases, the advantage of our proposed models becomes more significant compared to traditional methods, and remains competitive with regard to fully digital beamforming performance. During training, we also observed that smaller learning rates benefit larger arrays, likely due to increased sensitivity in optimization.

Fig. \ref{fig_multiantennas_runtime} further shows that the inference time of our three GNNs grows only slightly as the number of transmit antennas $N_t$ increases, while AMO and ICD increase exponentially (note the y-axis is in log scale). These trends align well with our theoretical analysis in (\ref{GNN_O}), (\ref{AMO_O}), and the $O(N_t^3)$ complexity described in \cite{ICD} for ICD, where the computational complexities of AMO and ICD grow cubically with $N_t$, while our GNNs grow quadratically. Although LCMLP-GNN benefits from a simpler structure as a linear-combination of MLPs output representation without a message passing structure, resulting in faster online inference, our GNNs achieve significantly better spectral efficiency, outperforming LCMLP-GNN by approximately $125\%$.



\begin{figure}[!t]
\centering
\includegraphics[width=0.7\linewidth]{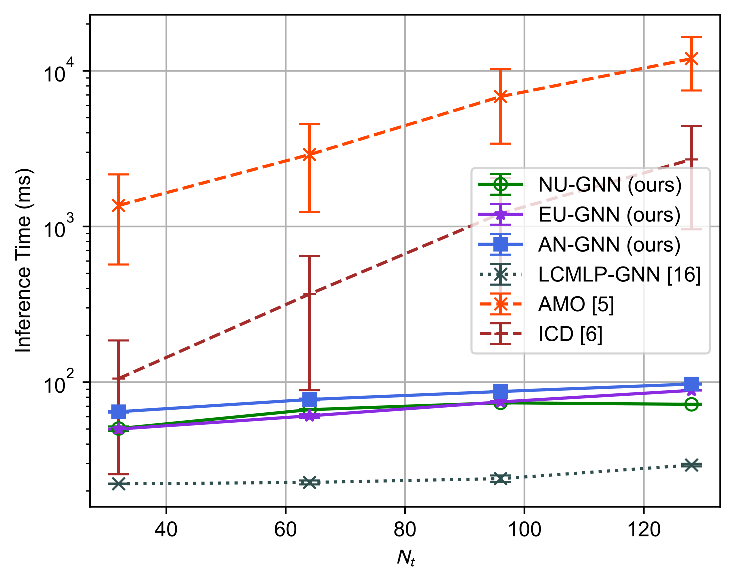}
\vspace{-5pt}
\caption{Inference running time comparison per CSI update on NVIDIA A100 GPU, averaged over $10^3$ CSI samples, versus the number of antennas $N_t$.}
\vspace{-15pt}
\label{fig_multiantennas_runtime}
\end{figure}

\begin{figure}[!t]
    \centering
    \subfloat[]{%
        \includegraphics[width=0.4\linewidth]{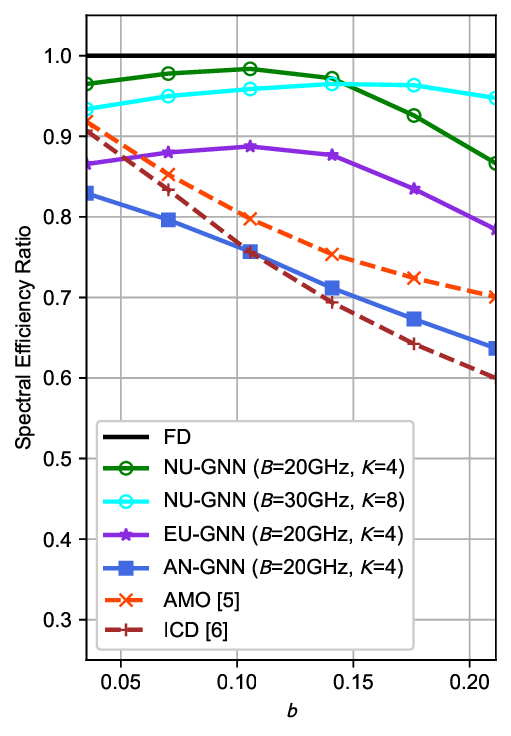}%
        \label{fig:beamsquint1}%
    }
    \hfil
    \subfloat[]{%
        \includegraphics[width=0.4\linewidth]{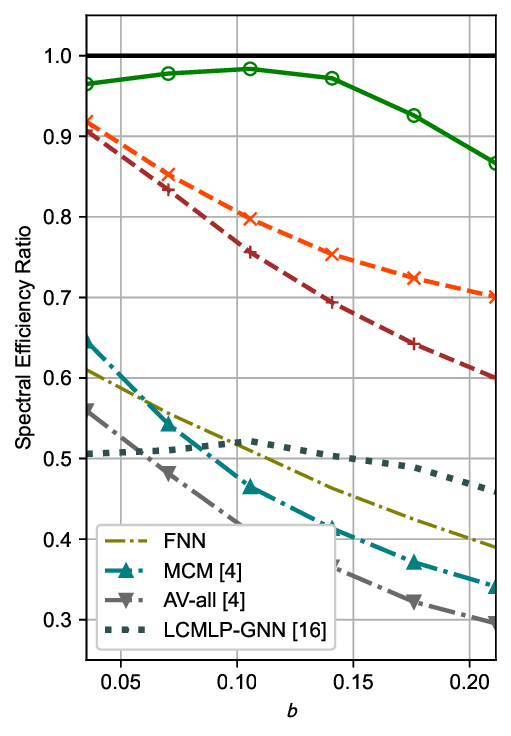}%
        \label{fig:beamsquint2}%
    }
    \caption{Spectral efficiency versus channel fractional bandwidth for different beamforming design algorithms, with $K=64$ subcarriers, $P_t=20$dBm, averaged over $10^3$ channel realizations. The central carrier frequency is $f_c = 142$GHz, and $b=\frac{B}{f_c}$, where $B$ is the communication channel bandwidth. The parameters in parentheses indicate the system settings used during training. (a) Comparison among the three proposed GNN structures, with an additional NU-GNN trained with $B$=30GHz, and $K=8$ subcarriers. (b) Comparison of NU-GNN with four other benchmarks.}
    \label{fig_beamsquint}
    \vspace{-2pt}
\end{figure}

\begin{figure}[!t]
    \centering
    \subfloat[]{%
        \raisebox{8pt}{\includegraphics[width=0.6\linewidth]{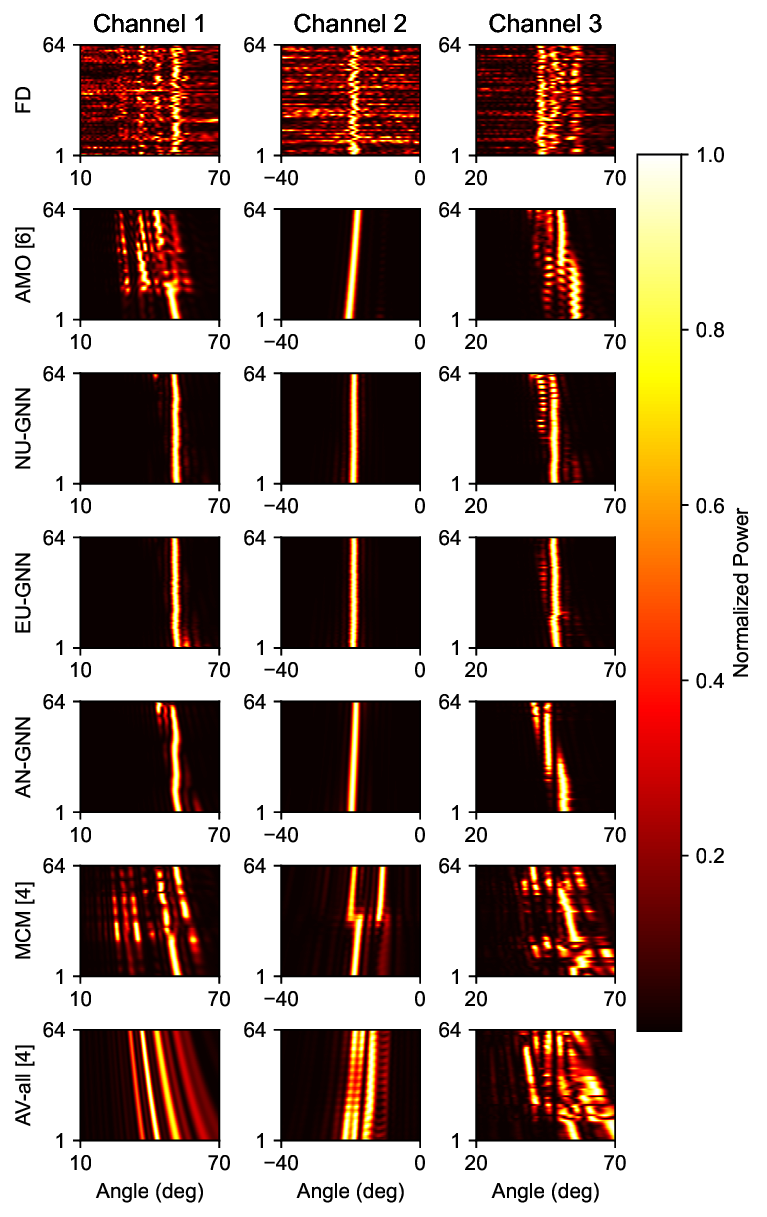}}%
        \label{fig:HeatMap}%
    }
    \subfloat[]{%
        \raisebox{2pt}{\includegraphics[width=0.367\linewidth]{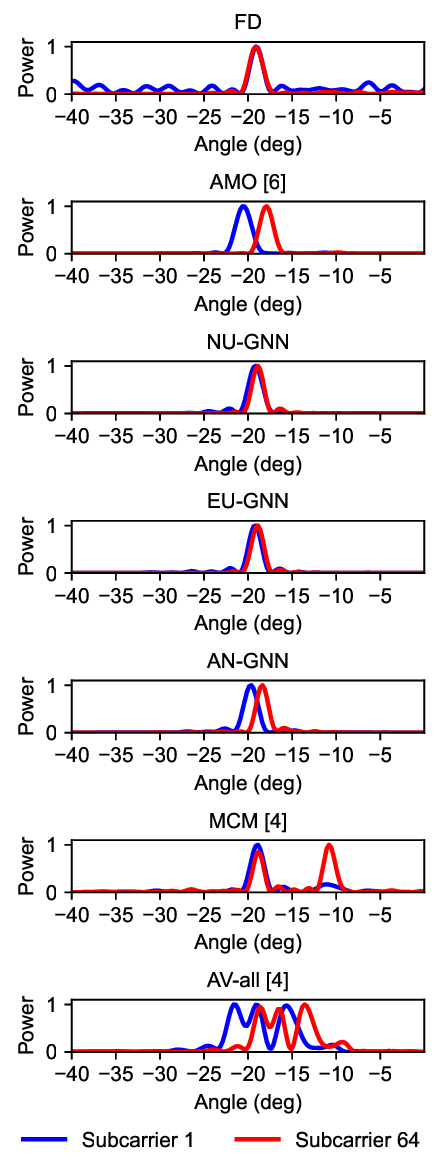}}%
        \label{fig:BeamPatterns}%
    }
    \caption{
    Visualization of beam squinting effects across three CSI samples. (a) Heatmap: Each row represents a beamforming method, with columns showing three distinct CSI samples. The x-axis is beam direction (degrees), and the y-axis is subcarrier index ($K=64$) for each subplot. Brighter colors indicate higher power gain. The slope of the bright region across subcarriers reflects the degree of beam squinting: a steep or vertical slope indicates less beam squint. (b) Beam patterns: The plot shows the beam patterns of the first and last subcarriers for a CSI sample. In the absence of beam squint, the two beam patterns should align perfectly.}
    \vspace{-2pt}
    \label{fig: heatmap}
\end{figure}

\textit{3) Beam Squinting Resiliency:} Fig. \ref{fig_beamsquint} shows that the proposed GNN structures effectively mitigate beam squinting. Let $b=\frac{B}{f_c}$ represent the fractional bandwidth. As $b$ increases, the beam squinting effect becomes pronounced as seen in all the baselines. Both our NU-GNN and EU-GNN, on the other hand, achieve a starkly superior resiliency to beam squinting than all baseline methods. Interestingly, both these GNN structures exhibit an optimal fractional bandwidth value for mitigating beam squinting. We further examine the effect of training bandwidths (at $B=20$GHz and $B=30$GHz) on the generalization ability of our GNN structures to larger bandwidths, and show that a larger training bandwidth leads to better generalization and strong resiliency to beam squinting. For AN-GNN, although it exhibits weaker beam squinting resiliency compared to NU-GNN and EU-GNN, it is still more resilient to beam squinting than the ICD algorithm \cite{ICD}.

The resiliency of our GNN models against beam squinting is further illustrated in Fig.~\ref{fig: heatmap}, showing the emitted power heatmap across subcarriers (left) and beam patterns at two representative subcarriers (right). In Fig. \ref{fig:HeatMap}, the slope of the bright regions reflects the degree of beam squinting: a steep or vertical slope indicates stronger resiliency (the same beam direction across all subcarriers), while the more tilted or curved patterns suggest the stronger beam squint. As shown, AMO exhibits noticeable directional shifts across all three channel samples, whereas our NU-GNN and EU-GNN maintain strong resiliency. Although AN-GNN exhibits minor shifts, it remains within an acceptable range compared to AMO. MCM and AV-all suffer from visible misalignment and display multiple side lobes across the subcarriers, leading to cluttered patterns that indicate unstable beam behavior and poor resiliency.

Fig. \ref{fig:BeamPatterns} compares the beam patterns of the first and last subcarriers for a CSI sample. In the absence of beam squint, the two beam patterns should align perfectly. The fully digital beamformer achieves this while AMO shows a $5^\circ$ shift, reflecting a moderate beam squinting effect. Both NU-GNN and EU-GNN exhibit near-perfect alignment, and AN-GNN shows only a minor deviation within $2^\circ$. In contrast, MCM and AV-all show distorted beam patterns with multiple side lobes and significant main lobe shifts. These irregularities suggest unfocused beam steering behavior and severe beam squinting.

\textit{4) Generalization over the Number of Subcarriers:} We tested the generalization ability of the proposed GNNs by varying the number of subcarriers and applying the trained MLPs to new subcarriers without retraining. As shown in Fig. \ref{fig_generalization}, even though we used $K=4$ subcarriers during the offline training process for all GNNs to save training time, in the online inference process, all three proposed GNN structures demonstrated excellent generalization ability to much larger $K$ values (up to $K=64$) without the need for retraining. Both the NU-GNN and EU-GNN structures maintain consistently high performance, on par with AMO, across all subcarrier configurations. Notably, for the AN-GNN structure, when comparing the blue and the light blue lines, the use of the attention mechanism in the aggregation operation (see Fig. \ref{fig_sa}) significantly outperforms the element-wise mean aggregation function by improving the GNN's generalization ability. This demonstrates the importance of including attention in the AN-GNN structure as it can dynamically assign a weight to signify the relative importance of each newly added subcarrier.

\begin{figure}[!t]
    \centering
    \subfloat[]{%
        \includegraphics[width=0.4\linewidth]{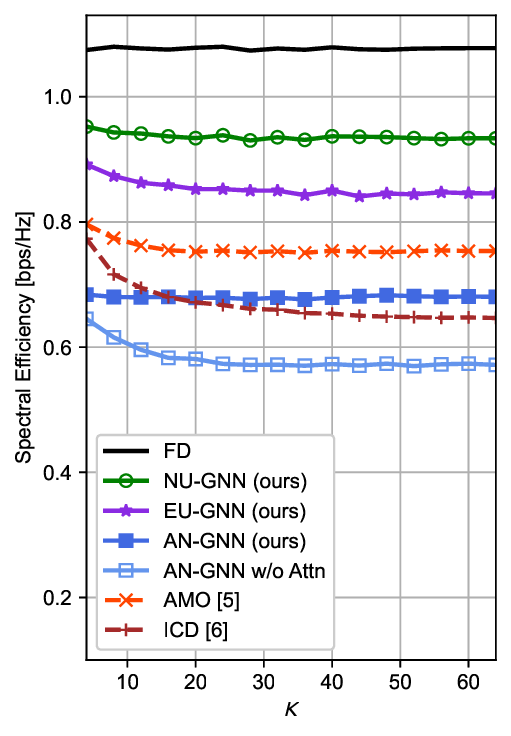}%
        \label{fig:generalization}%
    }
    \hfil
    \subfloat[]{%
        \includegraphics[width=0.4\linewidth]{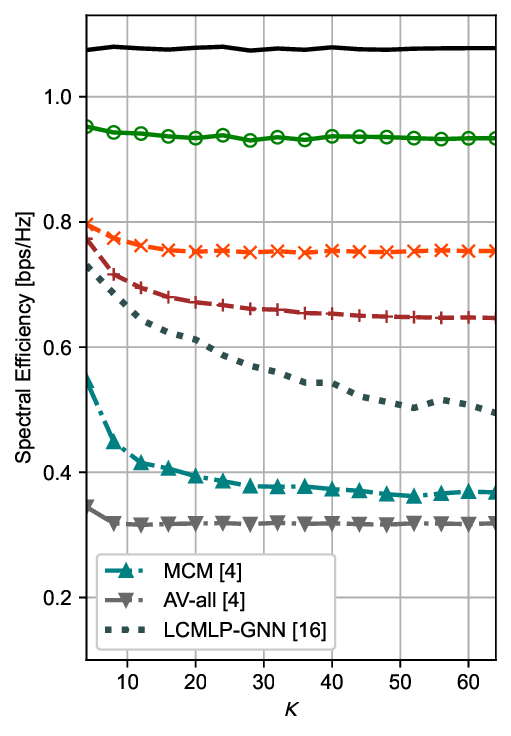}%
        \label{fig:generalization2}%
    }
    \caption{Spectral efficiency versus the number of subcarriers, when varying $K$ from 4 to 64, $P_t=20$dBm, averaged over $10^3$ channel realizations. The GNN models were trained with $K=4$ and then applied to all systems without retraining, with an additional AN-GNN trained without an attention-based aggregation. (a) Comparison among the three proposed GNN structures. (b) Comparison of NU-GNN with three other benchmarks.}
    \vspace{-7pt}
    \label{fig_generalization}
\end{figure}
\begin{figure}[!t]
\centering
\includegraphics[width=0.7\linewidth]{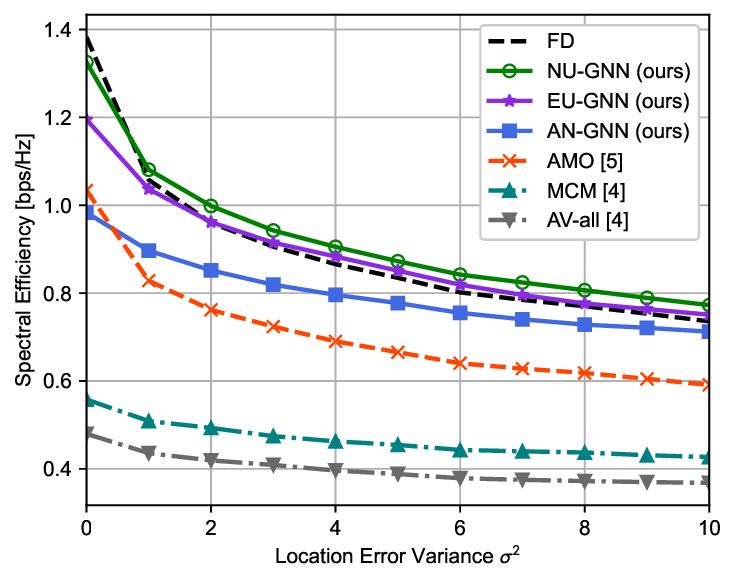}
\vspace{-5pt}
\caption{Spectral efficiency versus the location error variance $\sigma^2$, with $K=64$ subcarriers, $P_t=20$dBm, averaged over $10^3$ channel realizations.}
\vspace{-7pt}
\label{imperfect_CSI}
\end{figure}

\textit{5) Generalization over the Number of Users:}
\begin{figure}[!t]
\centering
\includegraphics[width=0.7\linewidth]{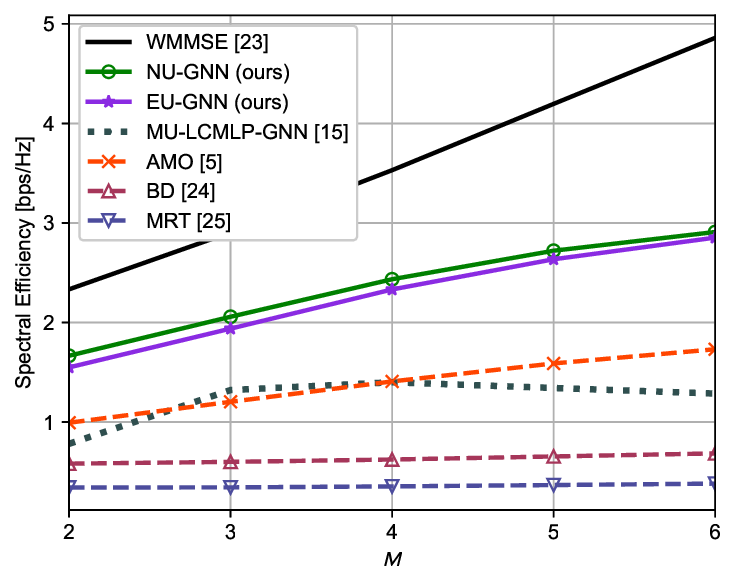}
\vspace{-5pt}
\caption{Spectral efficiency versus the number of users, with $K=64$ subcarriers, $P_t=20$dBm, averaged over $10^3$ channel realizations.}
\vspace{0pt}
\label{MU_gen}
\end{figure}

We further evaluate the generalization capability of the proposed GNN models in multi-user scenarios. The models are trained with $N_d=2$ data streams and $N_r=2$ receive antennas per user. The BS is equipped with $N_{\text{RF}}=12$ RF chains, $N_t=64$ transmit antennas, serving $M = 3$ users over $K = 4$ subcarriers. Both NU-GNN and EU-GNN have $L=3$ layers, which is chosen as it yields the best performance among the tested values $L=[2,3,4]$ (note that more GNN layers can lead to the over-smoothing effect and reduce learning). During training, weight decay regularization with a coefficient of $10^{-4}$ is employed to improve the performance. We then assess the generalization ability of the trained GNN models under different user configurations.

The baseline methods considered in this comparison are
\begin{itemize}[leftmargin=10pt]
    \item \textbf{WMMSE \cite{WMMSE}:} Fully digital beamforming method, which optimizes the beamformers by the weighted minimal mean-square error (WMMSE) algorithm, serving as a performance upper bound.
    \item \textbf{AMO \cite{AMO}:} Iteratively updating the analog beamformer via manifold optimization and alternating with digital beamforming update.
    \item \textbf{BD \cite{BD}:} Fully digital beamforming method, which applies block diagonalization to suppress inter-user interference.
    \item \textbf{MRT \cite{MRT}:} A fully digital beamforming method that maximizes the received signal power by aligning the precoder with the channel.
    \item \textbf{MU-LCMLP-GNN \cite{Wang2024GNN}:} An existing GNN method using a linear combination of the learned representation without message-passing for the multi-user scenario.
\end{itemize}

During the online inference process, all trained models are tested with $K=64$ subcarriers, which is much higher than the number of trained subcarriers $K=4$. As shown in Fig. \ref{MU_gen}, our proposed GNN models demonstrate strong generalization across different numbers of users and outperform AMO, MU-LCMLP-GNN, and the two fully digital beamforming methods, BD and MRT, across all user configurations. However, a performance gap remains between our proposed models and the fully digital WMMSE benchmark. We conjecture that this gap can be further shortened by designing new GNN architectures to explicitly model the inter-user interference. The exploration of this direction is left for future work.

\textit{6) Robustness to Imperfect CSI:} To further assess the robustness of our proposed methods under practical conditions, we evaluate their performance under imperfect CSI scenarios. Since we use the double-directional channel model in (\ref{channel}), we focus on the impact of UE location errors on channel CSI. Because this model relies on an accurate knowledge of the path angles, even small mismatches between the estimated and true UE locations can lead to significant CSI errors.

Denote the true UE location as $\mathbf{p}=[x,y]^T$, and the estimated UE location as $\hat{\mathbf{p}}=[\hat{x}, \hat{y}]^T$. The UE location error is
{\setlength{\abovedisplayskip}{-8pt}
 \setlength{\belowdisplayskip}{4pt}
\begin{align}
    \Delta \mathbf{p}=[\Delta x, \Delta y]^T &= [\hat{x}-x, \hat{y}-y]^T,
\end{align}
}which can be modeled as a two-dimensional Gaussian distribution centered at the true location:
{\setlength{\abovedisplayskip}{0pt}
 \setlength{\belowdisplayskip}{2pt}
\begin{align}
    \Delta\mathbf{p} \sim \mathcal{N}(0,\sigma^2\mathbf{I}_2).
\end{align}
}Here, the variance $\sigma^2$ controls the spread of the location error. In this evaluation, we use the imperfectly estimated channel $\hat{\mathbf{H}}(\hat{\mathbf{p}})$ to train the GNNs to learn the beamformers, but compute the spectral efficiency using the true channel $\mathbf{H}(\mathbf{p})$ to reflect realistic performance under imperfect CSI.

Fig. \ref{imperfect_CSI} shows the spectral efficiency versus the location error variance $\sigma^2$. As the location error increases, all methods show performance degradation. Notably, the more optimal a design, such as the fully digital beamforming method, the more sensitive it becomes to location errors, as it relies heavily on precise CSI for each subcarrier. In contrast, our proposed NU-GNN and EU-GNN maintain better resiliency and achieve stronger spectral efficiency than fully digital beamforming across the entire error range. Due to the data-driven nature of these learning-based methods, the GNNs are trained to capture statistical patterns rather than depend solely on exact CSI, giving them better generalization ability and making them inherently more robust to such imperfections compared to traditional methods like AMO, leading to higher performance.

\section{Conclusion}
We proposed three novel GNN structures for efficient hybrid beamforming design in multicarrier wideband MIMO systems, while effectively mitigating the beam squinting effect. By capturing the unique structure of hybrid beamforming in an OFDM system via a bipartite graph, we designed an efficient yet highly effective message-passing mechanism to optimize the GNN performance. The proposed GNNs not only optimize both digital and analog beamforming matrices but also adjust them dynamically to any change in the number of subcarriers by scaling the subcarrier nodes without the need for retraining.

Among the proposed 3 GNN structures, NU-GNN achieves the highest spectral efficiency performance with the lowest average inference time, while EU-GNN is the most stable during both training and inference. Interestingly, the hybrid AN-GNN structure exhibits significant computational savings during training because of fewer parameters, but this advantage vanishes at inference time because of the required computation for digital beamformers.

Comparing against traditional signal processing algorithms and existing GNN designs, our GNNs offer significant advantages in spectral efficiency, computational complexity, running time, and memory requirements, while achieving superior performance in beam squinting mitigation and robustness to imperfect CSI. These advantages make the proposed GNNs a viable solution for real-time beamforming adaptation. Finally, we demonstrated that the proposed NU-GNN and EU-GNN can be directly extended to multi-user scenarios, which outperform all existing ML multi-user solutions and exhibit strong generalization across users.

\appendices
\section*{Appendix A: Proof of Proposition 1}
Permutation equivariance is defined as
$f(\mathbf{\Pi}\mathbf{x}) = \mathbf{\Pi}f(\mathbf{x})$,
where $\mathbf{\Pi} \in \mathbb{R}^{N \times N}$ is a permutation matrix, $\mathbf{x}$ is the input, $f(\cdot)$ is a mapping function, and $f(\mathbf{x})$ denotes the corresponding output.
Here we employ
$\mathbf{\Pi}_{1} \in \mathbb{R}^{N_{\text{RF}} \times N_{\text{RF}}}$, $\mathbf{\Pi}_{2} \in \mathbb{R}^{N_s \times N_s}$, and $\mathbf{\Pi}_{3} \in \mathbb{R}^{K \times K}$ to represent permutations of the RF chain, data stream, and subcarrier indices, respectively. Since the digital beamformer $\mathbf{F} \in \mathbb{C}^{N_{\text{RF}}\times N_s \times K}$ involves permutations over multiple dimensions, we introduce a permutation operator as $[\pi(1),...,\pi(N)]=[1,...,N]\mathbf{\Pi}$. Then the permuted matrix multiplication is $\mathbf{H}'_ke^{j\mathbf{\Phi}_i'} \mathbf{F}'_{i,j,k}=\mathbf{H}_{\pi_3(k)}e^{j\mathbf{\Phi}_{\pi_1(i)}} \mathbf{F}_{\pi_1(i),\pi_2(j),\pi_3(k)}$, where $k=1,...,K$, $i=1,...,N_{\text{RF}}$ and $j=1,...,N_{s}$. Viewing the optimization problem (\ref{prob_psi}) as a solution mapping: $(\mathbf{\Phi}',\mathbf{F}')=\text{argmax}\ \mathcal{L}(\mathbf{H}')$, we have $(\mathbf{\Phi}_{\pi_1(i)},\mathbf{F}_{\pi_1(i),\pi_2(j),\pi_3(k)})=\text{argmax}\ \mathcal{L}(\mathbf{H}_{\pi_3(k)})$. And since $\mathbf{\Phi}'$ and $\mathbf{F}'$ still satisfy the power constraint in (\ref{prob_psi}), the optimization problem is permutation equivariant.

\section*{Appendix B: Proof of Proposition 2}
Let $\pi$ be any permutation over the subcarrier index $k$, and define the permutation operator as $\mathbf{b}_{\pi(k)}=\pi(\mathbf{b}_k)$. By applying the same permutation to both sides of Eq.~(\ref{GNN_mapping}), we obtain $\mathbf{b}_{\pi(k)}^{(l)}=\pi\left(f_{\mathbf{b}}(\mathbf{a}_k^{(l-1)},\phi(\cdot)\right)$. Since $f_{\mathbf{b}}(\cdot)$ is shared across all subcarriers $k$, it is independent of the subcarrier index. Furthermore, as all aggregation functions $\phi(\cdot)$ used in our design are permutation invariant, we have $\pi\left(f_{\mathbf{b}}\left(\mathbf{a}_k^{(l-1)},\phi(\cdot)\right)\right)=f_{\mathbf{b}}\left(\mathbf{a}_{\pi(k)}^{(l-1)}, \phi(\cdot)\right)$, which implies that $\mathbf{b}_{\pi(k)}^{(l)}=f_{\mathbf{b}}\left(\mathbf{a}_{\pi(k)}^{(l-1)}, \phi(\cdot)\right), \; \forall~\mathbf{a}_k, \mathbf{b}_k\in \mathcal{A}, \; \forall k$. At the final beamformer reconstruction step in Fig. \ref{fig:GNN_stru}, the learned representations $\mathbf{b}_{\pi(k)}^{(L)}$ are converted into beamforming matrices $\mathbf{\Phi}$ and $\mathbf{F}_{\pi(k)}$ accordingly, which naturally inherit permutation equivariance established by GNN updating layers.

\section*{Appendix C: Proof of Proposition 3}

The permutation equivariance with respect to subcarrier order has been established in Appendix B.
In the multi-user extension, the graph is constructed by grouping subcarrier nodes into user-specific subsets without modifying the GNN updating rules.
A permutation over user indices is therefore equivalent to a block-wise permutation over the corresponding subcarrier groups, which forms a subset of all possible subcarrier permutations.
Hence, the permutation equivariance property is preserved, and the extended graph remains permutation equivariant with respect to both subcarrier and user indices.

\end{document}